\documentclass[aps,prl,10pt,twocolumn,floatfix,superscriptaddress,showpacs]{revtex4-2}

\usepackage{amsmath, amssymb}
\usepackage{mathtools}
\usepackage{lipsum}
\usepackage{times}
\usepackage{graphicx}
\usepackage{wasysym}
\usepackage{bm}
\usepackage{hyperref}
\usepackage{xspace}
\usepackage{siunitx}
\usepackage{xcolor}
\usepackage{braket}
\usepackage{soul}

\definecolor{myColor}{rgb}{0.02,0.12,0.8}
\def\js#1{{\color{black} #1}}

\hypersetup{colorlinks=true, linkcolor=myColor, filecolor=myColor, urlcolor=myColor, citecolor=myColor, urlcolor=myColor}
\newcommand{\UD}{\ensuremath{U_{\text{D}}}}
\newcommand{\kD}{\ensuremath{k_{\text{D}}}}

\DeclareSIUnit{\nK}{\nano\kelvin}
\DeclareSIUnit{\aB}{\emph{a}_0}
\DeclareSIUnit{\G}{G}

\renewcommand{\figurename}[1]{Fig.~}
\newcommand{\kB}{k_{\text{B}}}

\newcommand{\Dc}{\mathcal{D}_{\text{c}}}
\newcommand{\Tc}{T_{\text{c}}}
\newcommand{\Nc}{N_{\text{c}}}

\begin{document}

\title{Observation of critical scaling in the Bose gas universality class}

\author{Leon Kleebank}
\affiliation{Institut f{\"u}r Angewandte Physik, Universit{\"a}t Bonn, Wegelerstra{\ss}e 8, 53115 Bonn, Germany}
\author{Frank Vewinger}
\affiliation{Institut f{\"u}r Angewandte Physik, Universit{\"a}t Bonn, Wegelerstra{\ss}e 8, 53115 Bonn, Germany}
\author{Arturo Camacho-Guardian}
\affiliation{Instituto de Física, Universidad Nacional Autónoma de México, Apdo. Postal 20-364, 01000 Cd. México, Mexico}
\author{Victor Romero-Rochín}
\affiliation{Instituto de Física, Universidad Nacional Autónoma de México, Apdo. Postal 20-364, 01000 Cd. México, Mexico}
\author{Rosario Paredes}
\affiliation{Instituto de Física, Universidad Nacional Autónoma de México, Apdo. Postal 20-364, 01000 Cd. México, Mexico}
\author{Martin Weitz}
\email[Corresponding author: ]{martin.weitz@uni-bonn.de}
\affiliation{Institut f{\"u}r Angewandte Physik, Universit{\"a}t Bonn, Wegelerstra{\ss}e 8, 53115 Bonn, Germany}
\author{Julian Schmitt}
\email[Corresponding author: ]{julian.schmitt@kip.uni-heidelberg.de}
\affiliation{Institut f{\"u}r Angewandte Physik, Universit{\"a}t Bonn, Wegelerstra{\ss}e 8, 53115 Bonn, Germany}
\affiliation{Kirchhoff-Institut f{\"u}r Physik, Universit{\"a}t Heidelberg, Im Neuenheimer Feld 225a, 69120 Heidelberg, Germany}

\begin{abstract}
Critical exponents characterize the divergent scaling of thermodynamic quantities near phase transitions and allow for the classification of physical systems into universality classes. While quantum gases thermalizing by interparticle interactions fall into the XY model universality class, the ideal Bose gas has been predicted to form a distinct universality class whose signatures have not yet been revealed experimentally. Here, we report the observation of critical scaling in a two-dimensional trapped quantum gas of essentially noninteracting photons, which thermalize by radiative contact to a reservoir of molecules inside a microcavity. By measuring the spatial correlations near the condensation transition, we determine the critical exponent for the correlation length to be $\nu = 0.52(4)$. Our results constitute a first experimental test of the long-standing scaling predictions for the Bose gas universality class.
\end{abstract}

\maketitle



Phase transitions play a pivotal role in nature, from quark-gluon plasmas inside nuclei, over superconductors in solid-state materials, to the formation of the early universe in cosmology~\cite{Rischke:2004,Vojta:2003,Guth:1980}. Near the phase transition at a critical temperature $\Tc$, where the thermodynamic state variables (e.g., pressure or density) change continuously upon varying the temperature $T$, the response of the system to external parameter variations (e.g., specific heat or compressibility) is drastically amplified. Large, correlated fluctuations of the order parameter occur, causing a divergence of the correlation length
\begin{equation}
    \xi\sim t^{-\nu}
    \label{eq:1}
\end{equation}
as a function of the reduced temperature $t = (T -\Tc)/\Tc$ with a certain critical exponent $\nu$. Such critical behavior is seen, for example, as critical opalescence in classical gas-liquid mixtures~\cite{Zubkov:1988}, or near the phase transition to superfluids in liquid helium~\cite{Lipa:2003,Pogorelov:2007}, ultracold atoms~\cite{Hung:2011,Ku:2012,Donner:2007,Navon:2015}, or exciton-polaritons~\cite{Alnatah:2024}.

As the correlation length $\xi$ diverges at the phase transition (i.e., at $t = 0$), intrinsic length scales of the system cease to be relevant. Phase transitions in this way reveal universal similarities between systems, which may be very different at the microscopic scale (e.g., superfluids and spin systems). Theoretically, it has been established that all systems within the same universality class display identical sets of critical exponents and scaling functions~\cite{Griffiths:1970,Kadanoff:1967,Fisher:1967}, with renormalization group methods providing their calculation from first principles~\cite{Wilson:1983}. Experimental studies have confirmed these predictions for interacting quantum many-body systems. For example, measurements of the critical exponent for the correlation length of $\nu= 0.67(13)$ in ultracold atomic gases near the Bose-Einstein condensation phase transition~\cite{Donner:2007,Navon:2015}, or determinations of the specific heat in superfluid helium and high-$\Tc$ superconductors~\cite{Lipa:2003,Pogorelov:2007,Overend:1994}, show good agreement with theory and firmly establish these systems as belonging to the three-dimensional (3D) XY universality class~\cite{Campostrini:2001}. Moreover, nonequilibrium universal behavior in interacting quantum gases has recently been observed using exciton-polaritons studying KPZ physics~\cite{Fontaine:2022, Widmann:2026}.

Remarkably, the ideal Bose gas of noninteracting particles, which was the system originally considered by Einstein in his proposal paper on Bose-Einstein condensation~\cite{Einstein:1925a}, forms a distinct universality class, characterized by critical exponents that differ from Landau theory~\cite{Gunton:1968,Schick:1969,Tarasov:2014}. In the homogeneous 3D gas, for instance, one expects an algebraic divergence of the correlation length at $\Tc$ with a critical exponent $\nu=1$~\cite{Kocharovsky:2010,Chatterjee:2014,Smith:2017,Reyes-Ayala:2019}, which differs from the (beyond-)mean-field results for interacting systems. Yet despite its fundamental importance, an experimental confirmation of these predictions has so far not been reported, largely because quantum gases are never perfectly noninteracting. As the low-temperature behavior of quantum gases depends sensitively on the density of states of low-lying excitations, even weak interactions modify density fluctuations near the critical point~\cite{Hadzibabic:2011}, causing these systems to fall into the XY rather than the Bose universality class. Furthermore, critical behavior also depends on dimensionality. In a homogeneous two-dimensional (2D) gas, where condensation is suppressed, only the interaction-driven Berezinskii-Kosterlitz-Thouless transition with an exponential divergence of $\xi$ is expected~\cite{Hadzibabic:2006,Jelic:2011}. Notably, 2D Bose-Einstein condensates have been observed in confined systems, specifically, harmonically trapped gases of atoms, exciton-polaritons, and photons~\cite{Fletcher:2015,Klaers:2010,Marelic:2015,Greveling:2018,Balili:2007}. Harmonic confinement however comes at the cost of an inhomogeneous density which limits systematic studies of $\xi$. The more recent realizations of Bose-Einstein condensates in box potentials have opened a promising route to controlled studies of critical phenomena in samples with uniform density~\cite{Navon:2015,Gaunt:2013,Busley:2022,Comaron:2025}. In contrast to atomic realizations, where interactions are required for thermalization, optical quantum gases allow one to reach the regime of nearly vanishing interactions.

Here we report the observation of critical behavior in a 2D photon gas near the phase transition to a Bose-Einstein condensate. The experiment is carried out in a box trap, where a phase transition in the thermodynamic limit is expected even for a gas with nearly uniform density, as well understood from a theoretical description using a Pöschl-Teller potential. Experimentally, the algebraic divergence of the correlation length obtained upon reaching the critical point provides a demonstration of critical scaling in the 2D Bose gas universality class. This physical setting is realized by harnessing a thermalization mechanism, which occurs by radiative contact to a molecule reservoir inside an optical cavity in the near absence of photon interactions. Our experimental result for the critical exponent for the correlation length of $\nu=0.52(4)$ is in good agreement with theoretical predictions.

\begin{figure}[t]
  \centering
  \includegraphics[width=1.0\columnwidth]{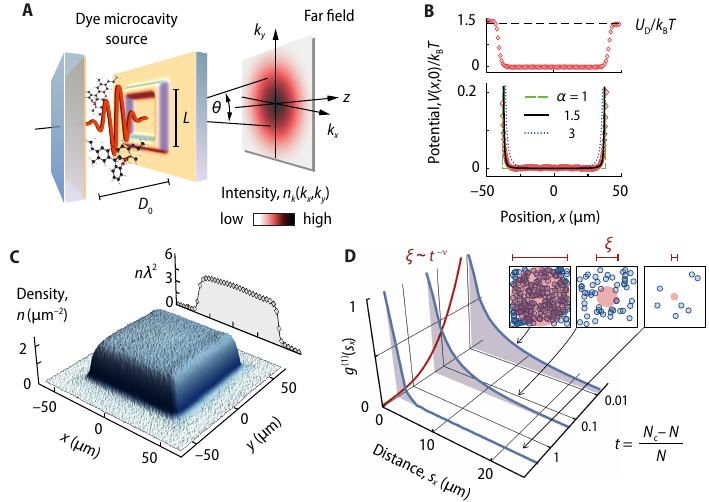}
 \caption{$\vert$ \textbf{Critical behavior in a uniform 2D Bose gas. (A)} Experimental system and measurement principle. Photons are trapped and thermalize in a dye-filled microcavity formed by a plane and a nanostructured mirror realizing a nearly uniform 2D Bose gas; the imprint on the mirror forms a confining square shaped box potential. The far-field angular intensity distribution, which corresponds to the photon momentum distribution $n_k(k_x,k_y)$, is related to $g^{(1)}(s_x,s_y)$ by a Fourier transform (see text). \textbf{(B)} Top panel: Horizontal cut through one of the experimentally used trap potentials with depth $U_\mathrm{D} = 1.4 \kB T$; the potential energy is calculated from the surface height profile of the mirror. Bottom panel: Zoom-in view on the data in the bottom part of the trap (symbols) shows the slightly softened edges as compared to a rigid box (dashed green line), well described by a Pöschl-Teller potential with $\alpha=1.5$ (solid black line). \textbf{(C)} Surface density of a quantum degenerate photon gas with phase-space density $n\lambda^2 = 4.8(5)$ in a box of $L = \SI{80}{\micro\meter}$. \textbf{(D)} First-order correlations $g^{(1)}(s_x)$, for increasing (decreasing) particle number $N$ (reduced temperature $t$). The correlation length $\xi$ (red line; see insets) grows algebraically with a critical exponent $\nu$ and diverges at the phase transition.}
\label{fig:1}
\end{figure}

Our 2D quantum gas of photons is created in a dye-filled optical microcavity, see Fig.~\ref{fig:1}(A), formed by two highly reflecting dielectric mirrors spaced by $D_0 \approx \SI{1.6}{\micro \meter}$, as described in earlier work~\cite{Busley:2022}. Due to the modified dispersion in the microcavity, the photons obtain an effective mass $m = 7.8\cdot 10^{-36}\SI{}{\kilo\gram}$ and can be trapped in square-shaped box potentials of variable size $L$ between roughly $\SI{40}{}$ and $\SI{80}{\micro\meter}$ by locally elevating the dielectric mirror coating using a direct laser writing technique~\cite{Kurtscheid:2020,Vretenar:2023}. The dye medium is pumped with a laser beam to inject excitations into the system. Consequently, the transverse cavity modes in the potential are populated by photons, which thermalize despite the near absence of interparticle interactions by radiative contact to dye molecules inside the optical cavity~\cite{Schmitt:2015}. Physically, the thermalization mechanism originates from the Kennard-Stepanov relation between the frequency-dependent absorption and emission rates of the dye molecules, $B_{\text{em}}(\omega)/B_{\text{abs}}(\omega)\sim \exp[-\hbar(\omega-\omega_{\text{zpl}})/\kB T]$ with zero-phonon frequency $\omega_{\text{zpl}}$, which results from fast (fs-scale) relaxation of the rovibrational molecular states in liquid solvents~\cite{Klaers:2010}. By absorption and re-emission processes the photon gas exchanges energy with the thermal molecule heat bath, leading to an equilibration of the photons at the dye rovibrational temperature $T = \SI{300}{\kelvin}$ and with a chemical potential $\mu \leq 0$ set by the excitation level of the dye~\cite{Schmitt:2018}. If thermalization occurs on a faster timescale than pumping and loss, the population of the cavity modes well corresponds to a Bose-Einstein distribution~\cite{Schmitt:2015,Kirton:2013}, so that critical behavior in the ideal Bose gas universality class is expected for the system despite its remaining driven-dissipative nature. Previous work has shown that an effective interaction strength in the photon gas would correspond to a dimensionless interaction parameter of $\widetilde{g}\approx 10^{-6}$~\cite{Dung:2017}, which is orders of magnitude smaller than in weakly-interacting Bose gases of atoms or exciton-polaritons~\cite{Hadzibabic:2006,Comaron:2025,Trypogeorgos:2025}. The spectral temperature yields a thermal wavelength $\lambda=(2\pi\hbar^2/m\kB T)^{1/2}=\SI{1.47}{\micro\meter}$ for the photons. The imprinted box trap, see Fig.~\ref{fig:1}(B), is well approximated at low energies by a potential that accounts for the experimentally rounded edges of the trap, $V(x,y) = \alpha(\alpha-1) \pi^2 \hbar^2/(2mL^2)\cdot [\cos^{-2}(\pi x/L)+ \cos^{-2}(\pi y/L)]$, known as a Pöschl-Teller potential~\cite{Poeschl:1933,Nieto:1978}. The dimensionless parameter $\alpha$ interpolates the Hamiltonian that describes the system between that of a photon gas in a rigid box ($\alpha=1$) and in a harmonic trap potential ($\alpha\rightarrow\infty$, with trap frequency $\alpha/L^2 = \mathrm{const.}$), respectively. For $\alpha\gtrsim 1$, as well fulfilled in the experiment [see Fig.~\ref{fig:1}(B)], bosons in this potential exhibit a phase transition to a Bose-Einstein condensate in the thermodynamic limit ($L\rightarrow\infty$ with $\alpha/L=\mathrm{const.}$, and keeping constant $T > 0$) at a finite temperature and critical phase-space density $\Dc=\Nc(\lambda/L)^2$ under near-homogeneous conditions (see SI). As shown in Fig.~\ref{fig:1}(C), the surface density $n(x,y)$ of the photon gas is visibly constant over essentially the entire bulk and rounded off slightly at the edges due to the box potential walls, so that to good approximation $n \approx N/L^2$. Our samples are quasi-homogeneous also at large phase-space densities $\mathcal{D}=n\lambda^2\gtrsim 1$, as seen in the exemplary data in Fig.~\ref{fig:1}(C), where the thermodynamics of the optical system are governed by quantum statistics, provided it remains below the critical value $\Dc$.

\begin{figure*}[t]
  \centering
  \includegraphics[width=0.8\textwidth]{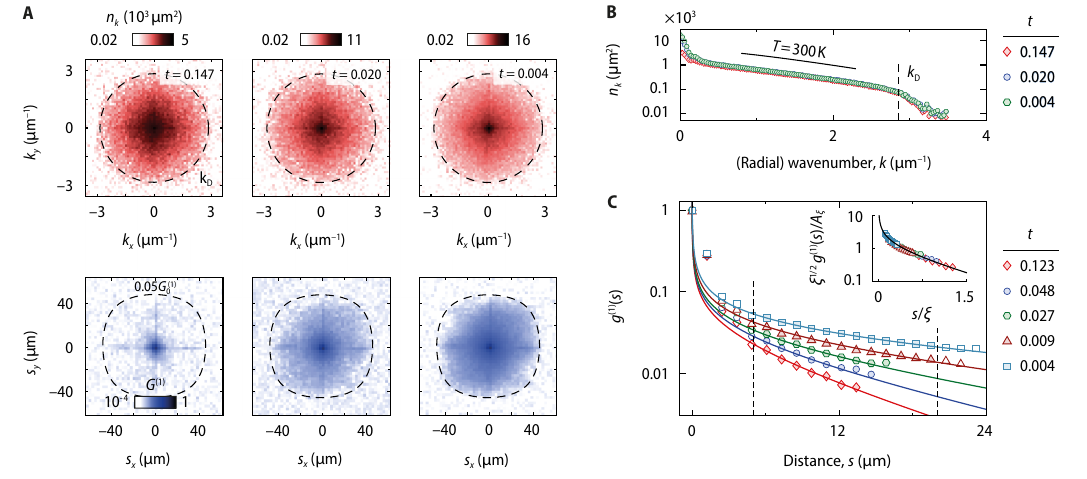}
 \caption{$\vert$ \textbf{Momentum distributions and growth of correlations. (A)} Far-field momentum distribution $n_k(k_x,k_y)$ (top, red) and derived coherence $G^{(1)}(s_x,s_y)$ (bottom, blue) at three different particle numbers, for $N = \{8.3, 9.4, 9.5\}\cdot 10^3$ and $L = \SI{60}{\micro\meter}$. The corresponding reduced temperatures $t$ are given in the top panels. Below $\Nc$ we observe short-range correlations corresponding to a Boltzmann distribution. As $N$ increases, the coherence grows due to the enhanced population at low $k$; dashed lines indicate trap depth wave number $\kD$ (top panel) and the maximum measurable correlation length, as the ground mode vanishes towards the box boundary (bottom), as discussed in the SI. All color scales are in logarithmic scale. \textbf{(B)} Radially averaged momentum distributions $n_k(k)$ for the quantum degenerate gas at the same values of $t$ as in (A) exhibit a thermal occupation of the excited states at $\SI{300}{\kelvin}$ (line, vertically shifted for better visibility). \textbf{(C)} Radially averaged first-order coherence $g^{(1)}(s)$ with fits (lines) in the region indicated by dashed lines and filled data points. The inset shows the correlations after rescaling with the fitted $\xi$, indicating the self-similarity of the correlations for the shown particle numbers.
 }
\label{fig:2}
\end{figure*}

To quantify the critical behavior of the photon gas close to the phase transition, we study the first-order correlation function. For a quantum-degenerate homogeneous Bose gas in $d$ dimensions~\cite{Fisher:1967},
\begin{equation}
    \langle \Psi^*({\mathbf{r}})\Psi({\mathbf{r'}})\rangle \sim s^{-(d-1)/2}e^{-s/\xi}
    \label{eq:2}
\end{equation}
describes the phase correlations of the bosonic field $\Psi({\mathbf{r}})$ between two points ${\mathbf{r}},{\mathbf{r'}}$ separated by $s=|\mathbf{r}-\mathbf{r'}|$. Figure~\ref{fig:1}(D) highlights the expected growth and divergence of the correlations when lowering the reduced temperature $t$ towards zero. A point to recall here is that Bose-Einstein condensation can be approached by increasing the particle number $N$, at constant temperature, towards the critical value $\Nc$ allowing for the equivalent expression for the reduced temperature as $t=(\Nc-N)/N$ for the system. At $t\gg 1$, the correlations are short-ranged and the correlation length $\xi$ on the order of the thermal wavelength $\lambda$. As $t$ is lowered, the wavelets overlap, leading to the growth and divergence of $\xi$. In the case of a 2D uniform gas [see Fig.~\ref{fig:1}(C)], the first-order spatial correlations between two points $(x,y)$ and $(x',y')$ are linked to the angular intensity distribution, i.e., to the momentum distribution $n_k(\mathbf{k})$, by
\begin{equation}
    G^{(1)}({\bf s}) = \langle \Psi^*(x,y) \Psi(x',y')\rangle \sim\int{n_k({\bf k}) e^{i {\bf k} \cdot {\bf s}} d^2{\bf k} },
    \label{eq:3}
\end{equation}
where $\mathbf{s}=(x-x',y-y')$ denotes the distance~\cite{Wolf:1975}. Note that eq.~\eqref{eq:3} has also been discussed in the context of ultracold atomic gases~\cite{Smith:2017} and we have verified its validity for thermalized photon gases in dye microcavity systems undergoing absorption-reemission processes (see SI).

Figure~\ref{fig:2}(A) shows 2D momentum distributions $n_k(k_x,k_y)$ of the optical quantum gas (top panels) in a $\SI{60}{\micro\meter}$-box for different reduced temperatures $t$. The signal is observed up to a wave number $k_\mathrm{D} = \sqrt{2 m \UD}/\hbar\approx \SI{2.9}{\per \micro\meter}$ determined by the trap depth $\UD = 1.4\kB T$~\cite{Busley:2022}. Above this energy, photons are no longer confined by the potential (see SI). As $t$ is reduced by increasing the photon number, the relative population at small momenta $k$ grows. In all panels, the measured distribution is radially symmetric within experimental uncertainties. The lower panels in Fig.~\ref{fig:2}(A) give $G^{(1)}(s_x,s_y)$ obtained by applying eq.~\eqref{eq:3} on the measured $n_k(k_x,k_y)$. Correspondingly, we find the spatial extent of the correlations to grow as the critical point at $t = 0$ is approached. The radially-averaged momentum distributions $n_k(k)$ are shown in Fig.~\ref{fig:2}(B) for the same values of $t$ as in Fig.~\ref{fig:2}(A); the agreement of the data points at $\SI{0.5}{\per\micro\meter}\lesssim k < \kD$ with the prediction for a Bose gas at $T = \SI{300}{\kelvin}$ (line, vertically shifted) confirms that the optical quantum gas in the cavity is well in thermal equilibrium with the dye solution at room temperature. Figure~\ref{fig:2}(C) shows corresponding radially averaged first-order coherence functions $g^{(1)}(s)$ at some more reduced temperatures $t$ for the same data set. Here we have introduced the normalized $g^{(1)}(s)$ to account for the finite size of the sample, which allows us to extract $\xi$ as described in the SI. Fitting the data with $g^{(1)}(s)= A_\xi s^{-1/2} \exp(-s/\xi)$ [see eq.~\eqref{eq:2} with $d = 2$] yields the correlation length $\xi$, which for the shown data grows from roughly $\SI{10}{\micro\meter}$ to $\SI{60}{\micro\meter}$. The rescaled plot in the inset of Fig.~\ref{fig:2}(C) shows that the correlation data is gradually shifted from an exponential (red data) towards a power-law (light blue) behavior as the critical phase-space density is approached, indicating the emergence of scale invariance in the limit of $\xi\rightarrow\infty$.

\begin{figure}[t]
  \centering
  \includegraphics[width=1.0\columnwidth]{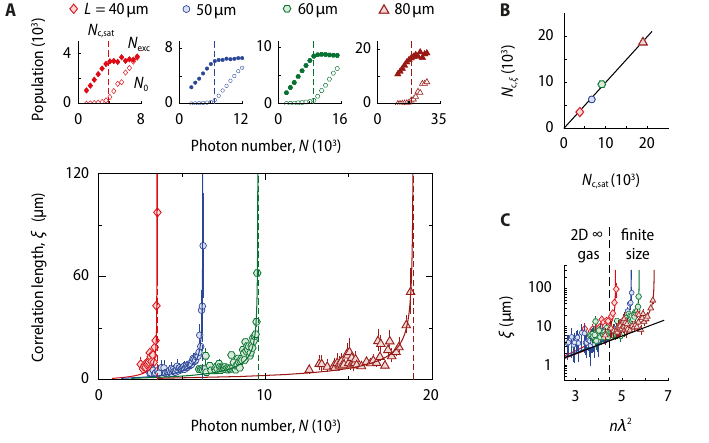}
 \caption{$\vert$ \textbf{Critical behavior at the condensation phase transition. (A)} Top panels: Occupation in ground (open symbols) and excited (filled symbols) states versus particle number for the investigated system sizes and $\alpha=\mathrm{const.}$; dashed lines indicate the critical points based on the saturation of the population in the excited states as determined by piecewise linear fits. Main panel: Measured correlation length $\xi$ along with theory prediction (lines), where $L$ is treated as a free parameter that is in good agreement with the imprinted structure sizes on the mirror. The critical particle numbers $N_{\mathrm{c},\xi} = \{3482, 6228, 9538, 18864\}$ are determined from the points of divergence (dashed lines). \textbf{(B)} The critical photon numbers extracted with the two methods based on the onset of saturation and the correlation length divergence agree within the error margins for all investigated system sizes. The solid curve with a unity slope is a guide to the eye. \textbf{(C)} Correlation length as a function of phase-space density for the four system sizes. The data follows the expected exponential growth (solid line) of the homogeneous 2D Bose gas (left of dashed line) until condensation sets in (right of dashed line). The error bars in all panels show combined systematic and statistical uncertainties.}
\label{fig:3}
\end{figure}

The observed growth of the correlation length when lowering the reduced temperature $t$ acts as a precursor for a phase transition in the photon gas. To systematically identify the critical point for the different studied system sizes, we have compared two signatures: the saturation of the excited mode population~\cite{Busley:2022,Tammuz:2011} and the correlation length as a function of the particle number $N$. Figure~\ref{fig:3}(A) shows the corresponding experimental data for different $L$. First, we extract the critical particle number $N_\mathrm{c,sat}$ by fits of the ground and excited state populations $N_0$ and $N_\mathrm{exc}=N-N_0$, respectively (top panels). We obtain $N_0$ by integrating the momentum space distribution $n_k(k_x,k_y)$ over a circular window of radius $\Delta k\approx4.4\pi/L$, centered at $(k_x,k_y)=(0,0)$, which essentially captures the ground state distribution~\cite{Busley:2022}. Secondly, we examine the extracted correlation length as a function of $N$ and observe a divergence at the critical photon number $N_{\mathrm{c},\xi}$ (main panel), where the correlations essentially span the entire system size. 

To determine $N_{\mathrm{c},\xi}$, the data is fitted by analytical theory curves for the Pöschl-Teller potential using $L$ as the only free parameter (solid lines), and the fit values for $L$ agree well with the independently determined system sizes from surface scans of our mirror samples (see SI). Figure~\ref{fig:3}(B) confirms that both critical particle numbers, $N_{\mathrm{c},\xi}$ and $N_{\mathrm{c,sat}}$, coincide within experimental uncertainties. The data shows that the critical particle numbers for larger system sizes increase, as understood from the requirement to reach the corresponding critical densities.

Additionally, our findings allow us to distinguish the physics of the infinite 2D homogeneous Bose gas, where true long-range order is suppressed, from the phase transition behavior within a soft-box potential. For this, we show the experimental data for the correlation length $\xi$ as a function of the phase-space density $n\lambda^2$ in Fig.~\ref{fig:3}(C). In the quantum-degenerate regime ($n\lambda^2>1$), $\xi$ first exhibits an exponential growth independent of the used system size $L$, in agreement with the theoretical prediction $\xi=\lambda/(4\pi)^{1/2} \exp(n\lambda^2/2)$ for the infinite system case~\cite{Hadzibabic:2011}. Near the critical phase-space density $\Dc$ that depends not only on the system size $L$ but also on the dimensionless trap parameter $\alpha$, however, the data starts to strongly deviate from the infinite-system scaling, entering a qualitatively different physical regime where the phase transition to a condensate starts to govern the (now divergent) scaling of $\xi$.

To make a prediction for the nature of the critical scaling behavior and its critical exponent $\nu$ (in the case of an algebraic divergence), we theoretically examine the dependence of the correlation length on the reduced temperature for $t\gtrsim 0$, as realized by particle numbers $N$ just below $\Nc$. Near the critical phase-space density $\Dc = \mathcal{D}(\mu=0)$ for Bose-Einstein condensation in the Pöschl-Teller potential, we obtain the approximate expression for the equation of state $\mathcal D(\mu)\approx \Dc - \mu/(\pi V_0)\ln(-\mu/\kB T)$, where $V_0 = \alpha^2\pi^2\hbar^2/(2mL^2)$. Using $t \simeq (\Dc-\mathcal{D})/\Dc$, one arrives at an explicit expression for the critical scaling close to the transition (see SI)
\begin{equation}
    \xi\approx \left[\frac{1}{2C_0} \frac{\ln(1/t)}{t}\right]^{1/2},
    \label{eq:4}
\end{equation}
with $C_0 = 2\pi^2 V_0 n_c/\kB T$, which indicates a critical exponent $\nu\approx 1/2$ with logarithmic corrections.

\begin{figure}[t]
  \centering
  \includegraphics[width=1.0\columnwidth]{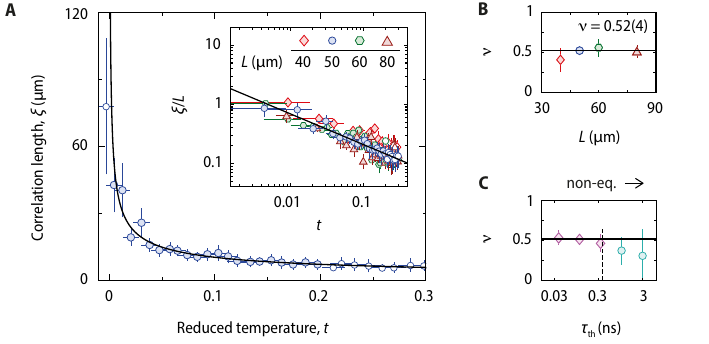}
 \caption{$\vert$ \textbf{Divergence of correlations and critical exponent. (A)} Correlation length $\xi$ as a function of reduced temperature $t$, exemplarily shown for $L = \SI{50}{\micro\meter}$ in the main panel. The solid line shows a fit (see main text), which yields $\nu= 0.52(5)$. Inset: Scaling the correlation length with the system size, all data sets collapse onto the same curve consistent with a power-law with exponent $\nu\approx 0.52$ (solid line), demonstrating the universal scaling behavior of the photon gas; a comparison with the interacting case is given in the SI. The filled symbols indicate the data used for fitting $\xi(t)$, while open symbols are not included. \textbf{(B)} The measured critical exponents $\nu$ for all system sizes give consistent results, averaging to $\nu=0.52(4)$ (solid line). \textbf{(C)} Critical exponent $\nu$ against thermalization time $\tau_\mathrm{th}$ for $L = \SI{50}{\micro\meter}$. For $\tau_\mathrm{th}$ exceeding the average photon cavity lifetime $\tau_\mathrm{cav}$ (vertical dashed line), the measured critical exponents (green circles) exhibit an apparent deviation from the equilibrium value (solid line). The error bars in all panels show combined systematic and statistical uncertainties.}
\label{fig:4}
\end{figure}

Figure~\ref{fig:4}(A) (main panel) shows the measured correlation length as a function of the reduced temperature $t = (N_{\mathrm{c},\xi}-N)/N$ calculated from the measured critical photon number $N_{\mathrm{c},\xi} = 6228$ for a system size $L = \SI{50}{\micro \meter}$ [see blue data in the main panel of Fig.~\ref{fig:3}(A)]. In absolute photon numbers, the data points range from $N = 4800$ to $N_{\mathrm{c},\xi}$ with a mean point spacing of $\Delta N = 50(6)$ photons, corresponding to a temperature resolution of $\Delta t = 0.010(2)$. We fit the data with an algebraically decaying function $\xi(t) = A_L t^{-\nu}$ (solid line) and obtain the critical exponent $\nu=0.52(5)$. We have carefully tested and confirmed the robustness of the fit result, by varying the range of fitted reduced temperatures $t$. For this we have taken into account both systematic and statistical uncertainties and excluded effects from data sampling; a detailed discussion of the total error budget is given in the SI. The logarithmic representation shown for all system sizes in the inset of Fig.~\ref{fig:4}(A) highlights the clear signature of a universal power-law scaling that is independent of the system size. The corresponding critical exponents $\nu$ are shown in Fig.~\ref{fig:4}(B). The obtained values are generally in good agreement with each other, yielding an average $\nu=0.52(4)$, which is our final result for the critical exponent.

Finally, we show the robustness of the critical exponent being a true equilibrium property. For this, we gradually remove the photon gas from equilibrium, as shown in Fig.~\ref{fig:4}(C). Experimentally, this is achieved by reducing the dye concentration in several steps from $\SI{3}{\milli\mol\per\liter}$ to $\SI{30}{\micro\mol\per\liter}$ and thereby extending the photon thermalization time from $\tau_\mathrm{th} = \SI{37}{\pico\second}$ to $\SI{3}{\nano\second}$ beyond the photon lifetime $\tau_\mathrm{cav}=\SI{350}{\pico\second}$ due to losses from mirror transmission (see SI); note that all previously shown experimental results were obtained for $\tau_\mathrm{th} = \SI{110}{\pico\second}$. For an incomplete thermalization of the photon gas, we observe that the density distribution remains uniform, but the momentum distribution $n_k(k)$ deviates from a thermal equilibrium distribution. Figure~\ref{fig:4}(C) summarizes the obtained values for $\nu$ extracted from $g^{(1)}(s)$, which agree with the equilibrium value (solid line) for short $\tau_\mathrm{th}$ and start to deviate substantially from it for $\tau_\mathrm{th}>\tau_\mathrm{cav}$. The extracted values for the critical exponents in the nonthermalized regime, however, are not physically meaningful, since the fit function assumes equilibrium conditions. The observed deviation indicates that the universal scaling behavior breaks down if the photon gas is not thermalized anymore. These results provide a further line of evidence that the measured critical exponent arises from the critical behavior near an equilibrium phase transition.

Our experiments establish the applicability of the concept of critical behavior near continuous phase transitions to quantum gases of light. By experimentally determining the algebraic scaling of the correlation length of an essentially noninteracting quantum gas trapped in a weakly confining Pöschl-Teller potential as the onset of Bose-Einstein condensation is approached, we demonstrate critical behavior for the ideal Bose gas universality class in two dimensions. While usually the critical exponents near phase transitions are governed by interactions~\cite{Lipa:2003,Hung:2011,Donner:2007,Navon:2015,Overend:1994} – which alter density fluctuations – in the here investigated photon gas full density-fluctuation-driven criticality is at work. This situation can be revealed because the window of reduced temperatures where interaction-induced criticality can be expected is extremely narrow, with $\Delta t = 10^{-7}$~\cite{Smith:2017}, which in our present, finite-size trapped system is inaccessible as it would imply changing the photon number by far below unity and correspondingly require much larger system sizes in the $\SI{}{\milli\meter}$-range. An experimental challenge for the future is to explore nonequilibrium universal phenomena, e.g., critical exceptional points~\cite{Hanai:2020,Zelle:2024}, or scaling behavior near non-thermal fixed points~\cite{Berges:2015b}, by studying the spatio-temporal correlations in homogeneous photon gases in the limit of incomplete thermalization.

\bibliography{references}

\begin{thebibliography}{55}%
\makeatletter
\providecommand \@ifxundefined [1]{%
 \@ifx{#1\undefined}
}%
\providecommand \@ifnum [1]{%
 \ifnum #1\expandafter \@firstoftwo
 \else \expandafter \@secondoftwo
 \fi
}%
\providecommand \@ifx [1]{%
 \ifx #1\expandafter \@firstoftwo
 \else \expandafter \@secondoftwo
 \fi
}%
\providecommand \natexlab [1]{#1}%
\providecommand \enquote  [1]{``#1''}%
\providecommand \bibnamefont  [1]{#1}%
\providecommand \bibfnamefont [1]{#1}%
\providecommand \citenamefont [1]{#1}%
\providecommand \href@noop [0]{\@secondoftwo}%
\providecommand \href [0]{\begingroup \@sanitize@url \@href}%
\providecommand \@href[1]{\@@startlink{#1}\@@href}%
\providecommand \@@href[1]{\endgroup#1\@@endlink}%
\providecommand \@sanitize@url [0]{\catcode `\\12\catcode `\$12\catcode
  `\&12\catcode `\#12\catcode `\^12\catcode `\_12\catcode `\%12\relax}%
\providecommand \@@startlink[1]{}%
\providecommand \@@endlink[0]{}%
\providecommand \url  [0]{\begingroup\@sanitize@url \@url }%
\providecommand \@url [1]{\endgroup\@href {#1}{\urlprefix }}%
\providecommand \urlprefix  [0]{URL }%
\providecommand \Eprint [0]{\href }%
\providecommand \doibase [0]{https://doi.org/}%
\providecommand \selectlanguage [0]{\@gobble}%
\providecommand \bibinfo  [0]{\@secondoftwo}%
\providecommand \bibfield  [0]{\@secondoftwo}%
\providecommand \translation [1]{[#1]}%
\providecommand \BibitemOpen [0]{}%
\providecommand \bibitemStop [0]{}%
\providecommand \bibitemNoStop [0]{.\EOS\space}%
\providecommand \EOS [0]{\spacefactor3000\relax}%
\providecommand \BibitemShut  [1]{\csname bibitem#1\endcsname}%
\let\auto@bib@innerbib\@empty
\bibitem [{\citenamefont {Rischke}(2004)}]{Rischke:2004}%
  \BibitemOpen
  \bibfield  {author} {\bibinfo {author} {\bibfnamefont {D.~H.}\ \bibnamefont
  {Rischke}},\ }\bibfield  {title} {\bibinfo {title} {The quark–gluon plasma
  in equilibrium},\ }\href
  {https://doi.org/https://doi.org/10.1016/j.ppnp.2003.09.002} {\bibfield
  {journal} {\bibinfo  {journal} {Prog. Part. Nucl. Phys.}\ }\textbf {\bibinfo
  {volume} {52}},\ \bibinfo {pages} {197} (\bibinfo {year} {2004})}\BibitemShut
  {NoStop}%
\bibitem [{\citenamefont {Vojta}(2003)}]{Vojta:2003}%
  \BibitemOpen
  \bibfield  {author} {\bibinfo {author} {\bibfnamefont {M.}~\bibnamefont
  {Vojta}},\ }\bibfield  {title} {\bibinfo {title} {Quantum phase
  transitions},\ }\href {https://doi.org/10.1088/0034-4885/66/12/R01}
  {\bibfield  {journal} {\bibinfo  {journal} {Rep. Prog. Phys.}\ }\textbf
  {\bibinfo {volume} {66}},\ \bibinfo {pages} {2069} (\bibinfo {year}
  {2003})}\BibitemShut {NoStop}%
\bibitem [{\citenamefont {Guth}\ and\ \citenamefont {Tye}(1980)}]{Guth:1980}%
  \BibitemOpen
  \bibfield  {author} {\bibinfo {author} {\bibfnamefont {A.~H.}\ \bibnamefont
  {Guth}}\ and\ \bibinfo {author} {\bibfnamefont {S.~H.~H.}\ \bibnamefont
  {Tye}},\ }\bibfield  {title} {\bibinfo {title} {Phase transitions and
  magnetic monopole production in the very early universe},\ }\href
  {https://doi.org/10.1103/PhysRevLett.44.631} {\bibfield  {journal} {\bibinfo
  {journal} {Phys. Rev. Lett.}\ }\textbf {\bibinfo {volume} {44}},\ \bibinfo
  {pages} {631} (\bibinfo {year} {1980})}\BibitemShut {NoStop}%
\bibitem [{\citenamefont {Zubkov}\ and\ \citenamefont
  {Romanov}(1988)}]{Zubkov:1988}%
  \BibitemOpen
  \bibfield  {author} {\bibinfo {author} {\bibfnamefont {L.~A.}\ \bibnamefont
  {Zubkov}}\ and\ \bibinfo {author} {\bibfnamefont {V.~P.}\ \bibnamefont
  {Romanov}},\ }\bibfield  {title} {\bibinfo {title} {Critical opalescence},\
  }\href {https://doi.org/10.1070/PU1988v031n04ABEH005749} {\bibfield
  {journal} {\bibinfo  {journal} {Sov. Phys. Uspekhi}\ }\textbf {\bibinfo
  {volume} {31}},\ \bibinfo {pages} {328} (\bibinfo {year} {1988})}\BibitemShut
  {NoStop}%
\bibitem [{\citenamefont {Lipa}\ \emph {et~al.}(2003)\citenamefont {Lipa},
  \citenamefont {Nissen}, \citenamefont {Stricker}, \citenamefont {Swanson},\
  and\ \citenamefont {Chui}}]{Lipa:2003}%
  \BibitemOpen
  \bibfield  {author} {\bibinfo {author} {\bibfnamefont {J.~A.}\ \bibnamefont
  {Lipa}}, \bibinfo {author} {\bibfnamefont {J.~A.}\ \bibnamefont {Nissen}},
  \bibinfo {author} {\bibfnamefont {D.~A.}\ \bibnamefont {Stricker}}, \bibinfo
  {author} {\bibfnamefont {D.~R.}\ \bibnamefont {Swanson}},\ and\ \bibinfo
  {author} {\bibfnamefont {T.~C.~P.}\ \bibnamefont {Chui}},\ }\bibfield
  {title} {\bibinfo {title} {Specific heat of liquid helium in zero gravity
  very near the lambda point},\ }\href
  {https://doi.org/10.1103/PhysRevB.68.174518} {\bibfield  {journal} {\bibinfo
  {journal} {Phys. Rev. B}\ }\textbf {\bibinfo {volume} {68}},\ \bibinfo
  {pages} {174518} (\bibinfo {year} {2003})}\BibitemShut {NoStop}%
\bibitem [{\citenamefont {Pogorelov}\ and\ \citenamefont
  {Suslov}(2007)}]{Pogorelov:2007}%
  \BibitemOpen
  \bibfield  {author} {\bibinfo {author} {\bibfnamefont {A.~A.}\ \bibnamefont
  {Pogorelov}}\ and\ \bibinfo {author} {\bibfnamefont {I.~M.}\ \bibnamefont
  {Suslov}},\ }\bibfield  {title} {\bibinfo {title} {On the critical exponents
  for the $\lambda$ transition in liquid helium},\ }\href
  {https://doi.org/10.1134/S0021364007130097} {\bibfield  {journal} {\bibinfo
  {journal} {JETP Lett.}\ }\textbf {\bibinfo {volume} {86}},\ \bibinfo {pages}
  {39} (\bibinfo {year} {2007})}\BibitemShut {NoStop}%
\bibitem [{\citenamefont {Hung}\ \emph {et~al.}(2011)\citenamefont {Hung},
  \citenamefont {Zhang}, \citenamefont {Gemelke},\ and\ \citenamefont
  {Chin}}]{Hung:2011}%
  \BibitemOpen
  \bibfield  {author} {\bibinfo {author} {\bibfnamefont {C.-L.}\ \bibnamefont
  {Hung}}, \bibinfo {author} {\bibfnamefont {X.}~\bibnamefont {Zhang}},
  \bibinfo {author} {\bibfnamefont {N.}~\bibnamefont {Gemelke}},\ and\ \bibinfo
  {author} {\bibfnamefont {C.}~\bibnamefont {Chin}},\ }\bibfield  {title}
  {\bibinfo {title} {Observation of scale invariance and universality in
  two-dimensional {B}ose gases},\ }\href {https://doi.org/10.1038/nature09722}
  {\bibfield  {journal} {\bibinfo  {journal} {Nature}\ }\textbf {\bibinfo
  {volume} {470}},\ \bibinfo {pages} {236} (\bibinfo {year}
  {2011})}\BibitemShut {NoStop}%
\bibitem [{\citenamefont {Ku}\ \emph {et~al.}(2012)\citenamefont {Ku},
  \citenamefont {Sommer}, \citenamefont {Cheuk},\ and\ \citenamefont
  {Zwierlein}}]{Ku:2012}%
  \BibitemOpen
  \bibfield  {author} {\bibinfo {author} {\bibfnamefont {M.~J.~H.}\
  \bibnamefont {Ku}}, \bibinfo {author} {\bibfnamefont {A.~T.}\ \bibnamefont
  {Sommer}}, \bibinfo {author} {\bibfnamefont {L.~W.}\ \bibnamefont {Cheuk}},\
  and\ \bibinfo {author} {\bibfnamefont {M.~W.}\ \bibnamefont {Zwierlein}},\
  }\bibfield  {title} {\bibinfo {title} {Revealing the superfluid lambda
  transition in the universal thermodynamics of a unitary {F}ermi gas},\ }\href
  {https://doi.org/10.1126/science.1214987} {\bibfield  {journal} {\bibinfo
  {journal} {Science}\ }\textbf {\bibinfo {volume} {335}},\ \bibinfo {pages}
  {563} (\bibinfo {year} {2012})}\BibitemShut {NoStop}%
\bibitem [{\citenamefont {Donner}\ \emph {et~al.}(2007)\citenamefont {Donner},
  \citenamefont {Ritter}, \citenamefont {Bourdel}, \citenamefont {Ottl},
  \citenamefont {K{\"o}hl},\ and\ \citenamefont {Esslinger}}]{Donner:2007}%
  \BibitemOpen
  \bibfield  {author} {\bibinfo {author} {\bibfnamefont {T.}~\bibnamefont
  {Donner}}, \bibinfo {author} {\bibfnamefont {S.}~\bibnamefont {Ritter}},
  \bibinfo {author} {\bibfnamefont {T.}~\bibnamefont {Bourdel}}, \bibinfo
  {author} {\bibfnamefont {A.}~\bibnamefont {Ottl}}, \bibinfo {author}
  {\bibfnamefont {M.}~\bibnamefont {K{\"o}hl}},\ and\ \bibinfo {author}
  {\bibfnamefont {T.}~\bibnamefont {Esslinger}},\ }\bibfield  {title} {\bibinfo
  {title} {Critical behavior of a trapped interacting {B}ose gas},\ }\href
  {https://doi.org/10.1126/science.1138807} {\bibfield  {journal} {\bibinfo
  {journal} {Science}\ }\textbf {\bibinfo {volume} {315}},\ \bibinfo {pages}
  {1556} (\bibinfo {year} {2007})}\BibitemShut {NoStop}%
\bibitem [{\citenamefont {Navon}\ \emph {et~al.}(2015)\citenamefont {Navon},
  \citenamefont {Gaunt}, \citenamefont {Smith},\ and\ \citenamefont
  {Hadzibabic}}]{Navon:2015}%
  \BibitemOpen
  \bibfield  {author} {\bibinfo {author} {\bibfnamefont {N.}~\bibnamefont
  {Navon}}, \bibinfo {author} {\bibfnamefont {A.~L.}\ \bibnamefont {Gaunt}},
  \bibinfo {author} {\bibfnamefont {R.~P.}\ \bibnamefont {Smith}},\ and\
  \bibinfo {author} {\bibfnamefont {Z.}~\bibnamefont {Hadzibabic}},\ }\bibfield
   {title} {\bibinfo {title} {Critical dynamics of spontaneous symmetry
  breaking in a homogeneous {B}ose gas},\ }\href
  {https://doi.org/10.1126/science.1258676} {\bibfield  {journal} {\bibinfo
  {journal} {Science}\ }\textbf {\bibinfo {volume} {347}},\ \bibinfo {pages}
  {167} (\bibinfo {year} {2015})}\BibitemShut {NoStop}%
\bibitem [{\citenamefont {Alnatah}\ \emph {et~al.}(2024)\citenamefont
  {Alnatah}, \citenamefont {Comaron}, \citenamefont {Mukherjee}, \citenamefont
  {Beaumariage}, \citenamefont {Pfeiffer}, \citenamefont {West}, \citenamefont
  {Baldwin}, \citenamefont {Szymańska},\ and\ \citenamefont
  {Snoke}}]{Alnatah:2024}%
  \BibitemOpen
  \bibfield  {author} {\bibinfo {author} {\bibfnamefont {H.}~\bibnamefont
  {Alnatah}}, \bibinfo {author} {\bibfnamefont {P.}~\bibnamefont {Comaron}},
  \bibinfo {author} {\bibfnamefont {S.}~\bibnamefont {Mukherjee}}, \bibinfo
  {author} {\bibfnamefont {J.}~\bibnamefont {Beaumariage}}, \bibinfo {author}
  {\bibfnamefont {L.~N.}\ \bibnamefont {Pfeiffer}}, \bibinfo {author}
  {\bibfnamefont {K.}~\bibnamefont {West}}, \bibinfo {author} {\bibfnamefont
  {K.}~\bibnamefont {Baldwin}}, \bibinfo {author} {\bibfnamefont
  {M.}~\bibnamefont {Szymańska}},\ and\ \bibinfo {author} {\bibfnamefont
  {D.~W.}\ \bibnamefont {Snoke}},\ }\bibfield  {title} {\bibinfo {title}
  {Critical fluctuations in a confined driven-dissipative quantum condensate},\
  }\href {https://doi.org/10.1126/sciadv.adi6762} {\bibfield  {journal}
  {\bibinfo  {journal} {Science Adv.}\ }\textbf {\bibinfo {volume} {10}},\
  \bibinfo {pages} {eadi6762} (\bibinfo {year} {2024})}\BibitemShut {NoStop}%
\bibitem [{\citenamefont {Griffiths}(1970)}]{Griffiths:1970}%
  \BibitemOpen
  \bibfield  {author} {\bibinfo {author} {\bibfnamefont {R.~B.}\ \bibnamefont
  {Griffiths}},\ }\bibfield  {title} {\bibinfo {title} {Dependence of critical
  indices on a parameter},\ }\href
  {https://doi.org/10.1103/PhysRevLett.24.1479} {\bibfield  {journal} {\bibinfo
   {journal} {Phys. Rev. Lett.}\ }\textbf {\bibinfo {volume} {24}},\ \bibinfo
  {pages} {1479} (\bibinfo {year} {1970})}\BibitemShut {NoStop}%
\bibitem [{\citenamefont {Kadanoff}\ \emph {et~al.}(1967)\citenamefont
  {Kadanoff}, \citenamefont {G\"otze}, \citenamefont {Hamblen}, \citenamefont
  {Hecht}, \citenamefont {Lewis}, \citenamefont {Palciauskas}, \citenamefont
  {Rayl}, \citenamefont {Swift}, \citenamefont {Aspnes},\ and\ \citenamefont
  {Kane}}]{Kadanoff:1967}%
  \BibitemOpen
  \bibfield  {author} {\bibinfo {author} {\bibfnamefont {L.~P.}\ \bibnamefont
  {Kadanoff}}, \bibinfo {author} {\bibfnamefont {W.}~\bibnamefont {G\"otze}},
  \bibinfo {author} {\bibfnamefont {D.}~\bibnamefont {Hamblen}}, \bibinfo
  {author} {\bibfnamefont {R.}~\bibnamefont {Hecht}}, \bibinfo {author}
  {\bibfnamefont {E.~A.~S.}\ \bibnamefont {Lewis}}, \bibinfo {author}
  {\bibfnamefont {V.~V.}\ \bibnamefont {Palciauskas}}, \bibinfo {author}
  {\bibfnamefont {M.}~\bibnamefont {Rayl}}, \bibinfo {author} {\bibfnamefont
  {J.}~\bibnamefont {Swift}}, \bibinfo {author} {\bibfnamefont
  {D.}~\bibnamefont {Aspnes}},\ and\ \bibinfo {author} {\bibfnamefont
  {J.}~\bibnamefont {Kane}},\ }\bibfield  {title} {\bibinfo {title} {Static
  phenomena near critical points: Theory and experiment},\ }\href
  {https://doi.org/10.1103/RevModPhys.39.395} {\bibfield  {journal} {\bibinfo
  {journal} {Rev. Mod. Phys.}\ }\textbf {\bibinfo {volume} {39}},\ \bibinfo
  {pages} {395} (\bibinfo {year} {1967})}\BibitemShut {NoStop}%
\bibitem [{\citenamefont {Fisher}(1967)}]{Fisher:1967}%
  \BibitemOpen
  \bibfield  {author} {\bibinfo {author} {\bibfnamefont {M.~E.}\ \bibnamefont
  {Fisher}},\ }\bibfield  {title} {\bibinfo {title} {The theory of equilibrium
  critical phenomena},\ }\href {https://doi.org/10.1088/0034-4885/30/2/306}
  {\bibfield  {journal} {\bibinfo  {journal} {Rep. Prog. Phys.}\ }\textbf
  {\bibinfo {volume} {30}},\ \bibinfo {pages} {615} (\bibinfo {year}
  {1967})}\BibitemShut {NoStop}%
\bibitem [{\citenamefont {Wilson}(1983)}]{Wilson:1983}%
  \BibitemOpen
  \bibfield  {author} {\bibinfo {author} {\bibfnamefont {K.~G.}\ \bibnamefont
  {Wilson}},\ }\bibfield  {title} {\bibinfo {title} {The renormalization group
  and critical phenomena},\ }\href {https://doi.org/10.1103/RevModPhys.55.583}
  {\bibfield  {journal} {\bibinfo  {journal} {Rev. Mod. Phys.}\ }\textbf
  {\bibinfo {volume} {55}},\ \bibinfo {pages} {583} (\bibinfo {year}
  {1983})}\BibitemShut {NoStop}%
\bibitem [{\citenamefont {Overend}\ \emph {et~al.}(1994)\citenamefont
  {Overend}, \citenamefont {Howson},\ and\ \citenamefont
  {Lawrie}}]{Overend:1994}%
  \BibitemOpen
  \bibfield  {author} {\bibinfo {author} {\bibfnamefont {N.}~\bibnamefont
  {Overend}}, \bibinfo {author} {\bibfnamefont {M.~A.}\ \bibnamefont
  {Howson}},\ and\ \bibinfo {author} {\bibfnamefont {I.~D.}\ \bibnamefont
  {Lawrie}},\ }\bibfield  {title} {\bibinfo {title} {3{D} {X}-{Y} scaling of
  the specific heat of
  {${\mathrm{YBa}}_{2}$${\mathrm{Cu}}_{3}$${\mathrm{O}}_{7\mathrm{\ensuremath{-}}\mathrm{\ensuremath{\delta}}}$}
  single crystals},\ }\href {https://doi.org/10.1103/PhysRevLett.72.3238}
  {\bibfield  {journal} {\bibinfo  {journal} {Phys. Rev. Lett.}\ }\textbf
  {\bibinfo {volume} {72}},\ \bibinfo {pages} {3238} (\bibinfo {year}
  {1994})}\BibitemShut {NoStop}%
\bibitem [{\citenamefont {Campostrini}\ \emph {et~al.}(2001)\citenamefont
  {Campostrini}, \citenamefont {Hasenbusch}, \citenamefont {Pelissetto},
  \citenamefont {Rossi},\ and\ \citenamefont {Vicari}}]{Campostrini:2001}%
  \BibitemOpen
  \bibfield  {author} {\bibinfo {author} {\bibfnamefont {M.}~\bibnamefont
  {Campostrini}}, \bibinfo {author} {\bibfnamefont {M.}~\bibnamefont
  {Hasenbusch}}, \bibinfo {author} {\bibfnamefont {A.}~\bibnamefont
  {Pelissetto}}, \bibinfo {author} {\bibfnamefont {P.}~\bibnamefont {Rossi}},\
  and\ \bibinfo {author} {\bibfnamefont {E.}~\bibnamefont {Vicari}},\
  }\bibfield  {title} {\bibinfo {title} {Critical behavior of the
  three-dimensional $\mathrm{XY}$ universality class},\ }\href
  {https://doi.org/10.1103/PhysRevB.63.214503} {\bibfield  {journal} {\bibinfo
  {journal} {Phys. Rev. B}\ }\textbf {\bibinfo {volume} {63}},\ \bibinfo
  {pages} {214503} (\bibinfo {year} {2001})}\BibitemShut {NoStop}%
\bibitem [{\citenamefont {Fontaine}\ \emph {et~al.}(2022)\citenamefont
  {Fontaine}, \citenamefont {Squizzato}, \citenamefont {Baboux}, \citenamefont
  {Amelio}, \citenamefont {Lema{\^i}tre}, \citenamefont {Morassi},
  \citenamefont {Sagnes}, \citenamefont {Le~Gratiet}, \citenamefont {Harouri},
  \citenamefont {Wouters}, \citenamefont {Carusotto}, \citenamefont {Amo},
  \citenamefont {Richard}, \citenamefont {Minguzzi}, \citenamefont {Canet},
  \citenamefont {Ravets},\ and\ \citenamefont {Bloch}}]{Fontaine:2022}%
  \BibitemOpen
  \bibfield  {author} {\bibinfo {author} {\bibfnamefont {Q.}~\bibnamefont
  {Fontaine}}, \bibinfo {author} {\bibfnamefont {D.}~\bibnamefont {Squizzato}},
  \bibinfo {author} {\bibfnamefont {F.}~\bibnamefont {Baboux}}, \bibinfo
  {author} {\bibfnamefont {I.}~\bibnamefont {Amelio}}, \bibinfo {author}
  {\bibfnamefont {A.}~\bibnamefont {Lema{\^i}tre}}, \bibinfo {author}
  {\bibfnamefont {M.}~\bibnamefont {Morassi}}, \bibinfo {author} {\bibfnamefont
  {I.}~\bibnamefont {Sagnes}}, \bibinfo {author} {\bibfnamefont
  {L.}~\bibnamefont {Le~Gratiet}}, \bibinfo {author} {\bibfnamefont
  {A.}~\bibnamefont {Harouri}}, \bibinfo {author} {\bibfnamefont
  {M.}~\bibnamefont {Wouters}}, \bibinfo {author} {\bibfnamefont
  {I.}~\bibnamefont {Carusotto}}, \bibinfo {author} {\bibfnamefont
  {A.}~\bibnamefont {Amo}}, \bibinfo {author} {\bibfnamefont {M.}~\bibnamefont
  {Richard}}, \bibinfo {author} {\bibfnamefont {A.}~\bibnamefont {Minguzzi}},
  \bibinfo {author} {\bibfnamefont {L.}~\bibnamefont {Canet}}, \bibinfo
  {author} {\bibfnamefont {S.}~\bibnamefont {Ravets}},\ and\ \bibinfo {author}
  {\bibfnamefont {J.}~\bibnamefont {Bloch}},\ }\bibfield  {title} {\bibinfo
  {title} {{K}ardar--{P}arisi--{Z}hang universality in a one-dimensional
  polariton condensate},\ }\href {https://doi.org/10.1038/s41586-022-05001-8}
  {\bibfield  {journal} {\bibinfo  {journal} {Nature}\ }\textbf {\bibinfo
  {volume} {608}},\ \bibinfo {pages} {687} (\bibinfo {year}
  {2022})}\BibitemShut {NoStop}%
\bibitem [{\citenamefont {Widmann}\ \emph {et~al.}(2026)\citenamefont
  {Widmann}, \citenamefont {Dam}, \citenamefont {Düreth}, \citenamefont
  {Mayer}, \citenamefont {Daviet}, \citenamefont {Zelle}, \citenamefont
  {Laibacher}, \citenamefont {Emmerling}, \citenamefont {Kamp}, \citenamefont
  {Diehl}, \citenamefont {Betzold}, \citenamefont {Klembt},\ and\ \citenamefont
  {Höfling}}]{Widmann:2026}%
  \BibitemOpen
  \bibfield  {author} {\bibinfo {author} {\bibfnamefont {S.}~\bibnamefont
  {Widmann}}, \bibinfo {author} {\bibfnamefont {S.}~\bibnamefont {Dam}},
  \bibinfo {author} {\bibfnamefont {J.}~\bibnamefont {Düreth}}, \bibinfo
  {author} {\bibfnamefont {C.~G.}\ \bibnamefont {Mayer}}, \bibinfo {author}
  {\bibfnamefont {R.}~\bibnamefont {Daviet}}, \bibinfo {author} {\bibfnamefont
  {C.~P.}\ \bibnamefont {Zelle}}, \bibinfo {author} {\bibfnamefont
  {D.}~\bibnamefont {Laibacher}}, \bibinfo {author} {\bibfnamefont
  {M.}~\bibnamefont {Emmerling}}, \bibinfo {author} {\bibfnamefont
  {M.}~\bibnamefont {Kamp}}, \bibinfo {author} {\bibfnamefont {S.}~\bibnamefont
  {Diehl}}, \bibinfo {author} {\bibfnamefont {S.}~\bibnamefont {Betzold}},
  \bibinfo {author} {\bibfnamefont {S.}~\bibnamefont {Klembt}},\ and\ \bibinfo
  {author} {\bibfnamefont {S.}~\bibnamefont {Höfling}},\ }\bibfield  {title}
  {\bibinfo {title} {Observation of kardar-parisi-zhang universal scaling in
  two dimensions},\ }\href {https://doi.org/10.1126/science.aeb4154} {\bibfield
   {journal} {\bibinfo  {journal} {Science}\ }\textbf {\bibinfo {volume}
  {392}},\ \bibinfo {pages} {221} (\bibinfo {year} {2026})},\ \Eprint
  {https://arxiv.org/abs/https://www.science.org/doi/pdf/10.1126/science.aeb4154}
  {https://www.science.org/doi/pdf/10.1126/science.aeb4154} \BibitemShut
  {NoStop}%
\bibitem [{\citenamefont {{E}instein}(1925)}]{Einstein:1925a}%
  \BibitemOpen
  \bibfield  {author} {\bibinfo {author} {\bibfnamefont {A.}~\bibnamefont
  {{E}instein}},\ }\bibfield  {title} {\bibinfo {title} {Quantentheorie des
  einatomigen idealen {G}ases. {Z}weite {A}bhandlung},\ }\href
  {https://doi.org/10.1002/3527608958.ch28} {\bibfield  {journal} {\bibinfo
  {journal} {Sitzungsber. d. Königl. Preuss. Akad. d. Wiss.}\ }\textbf
  {\bibinfo {volume} {1}},\ \bibinfo {pages} {3} (\bibinfo {year}
  {1925})}\BibitemShut {NoStop}%
\bibitem [{\citenamefont {Gunton}\ and\ \citenamefont
  {Buckingham}(1968)}]{Gunton:1968}%
  \BibitemOpen
  \bibfield  {author} {\bibinfo {author} {\bibfnamefont {J.~D.}\ \bibnamefont
  {Gunton}}\ and\ \bibinfo {author} {\bibfnamefont {M.~J.}\ \bibnamefont
  {Buckingham}},\ }\bibfield  {title} {\bibinfo {title} {Condensation of the
  ideal {B}ose gas as a cooperative transition},\ }\href
  {https://doi.org/10.1103/PhysRev.166.152} {\bibfield  {journal} {\bibinfo
  {journal} {Phys. Rev.}\ }\textbf {\bibinfo {volume} {166}},\ \bibinfo {pages}
  {152} (\bibinfo {year} {1968})}\BibitemShut {NoStop}%
\bibitem [{\citenamefont {Schick}\ and\ \citenamefont
  {Zilsel}(1969)}]{Schick:1969}%
  \BibitemOpen
  \bibfield  {author} {\bibinfo {author} {\bibfnamefont {M.}~\bibnamefont
  {Schick}}\ and\ \bibinfo {author} {\bibfnamefont {P.~R.}\ \bibnamefont
  {Zilsel}},\ }\bibfield  {title} {\bibinfo {title} {Order parameter,
  mean-field theory, and the ideal {B}ose gas},\ }\href
  {https://doi.org/10.1103/PhysRev.188.522} {\bibfield  {journal} {\bibinfo
  {journal} {Phys. Rev.}\ }\textbf {\bibinfo {volume} {188}},\ \bibinfo {pages}
  {522} (\bibinfo {year} {1969})}\BibitemShut {NoStop}%
\bibitem [{\citenamefont {Tarasov}\ \emph {et~al.}(2014)\citenamefont
  {Tarasov}, \citenamefont {Kocharovsky},\ and\ \citenamefont
  {Kocharovsky}}]{Tarasov:2014}%
  \BibitemOpen
  \bibfield  {author} {\bibinfo {author} {\bibfnamefont {S.~V.}\ \bibnamefont
  {Tarasov}}, \bibinfo {author} {\bibfnamefont {V.~V.}\ \bibnamefont
  {Kocharovsky}},\ and\ \bibinfo {author} {\bibfnamefont {V.~V.}\ \bibnamefont
  {Kocharovsky}},\ }\bibfield  {title} {\bibinfo {title} {Universal scaling in
  the statistics and thermodynamics of a {B}ose-{E}instein condensation of an
  ideal gas in an arbitrary trap},\ }\href
  {https://doi.org/10.1103/PhysRevA.90.033605} {\bibfield  {journal} {\bibinfo
  {journal} {Phys. Rev. A}\ }\textbf {\bibinfo {volume} {90}},\ \bibinfo
  {pages} {033605} (\bibinfo {year} {2014})}\BibitemShut {NoStop}%
\bibitem [{\citenamefont {Kocharovsky}\ and\ \citenamefont
  {Kocharovsky}(2010)}]{Kocharovsky:2010}%
  \BibitemOpen
  \bibfield  {author} {\bibinfo {author} {\bibfnamefont {V.~V.}\ \bibnamefont
  {Kocharovsky}}\ and\ \bibinfo {author} {\bibfnamefont {V.~V.}\ \bibnamefont
  {Kocharovsky}},\ }\bibfield  {title} {\bibinfo {title} {Analytical theory of
  mesoscopic {B}ose-{E}instein condensation in an ideal gas},\ }\href
  {https://doi.org/10.1103/PhysRevA.81.033615} {\bibfield  {journal} {\bibinfo
  {journal} {Phys. Rev. A}\ }\textbf {\bibinfo {volume} {81}},\ \bibinfo
  {pages} {033615} (\bibinfo {year} {2010})}\BibitemShut {NoStop}%
\bibitem [{\citenamefont {Chatterjee}\ and\ \citenamefont
  {Diaconis}(2014)}]{Chatterjee:2014}%
  \BibitemOpen
  \bibfield  {author} {\bibinfo {author} {\bibfnamefont {S.}~\bibnamefont
  {Chatterjee}}\ and\ \bibinfo {author} {\bibfnamefont {P.}~\bibnamefont
  {Diaconis}},\ }\bibfield  {title} {\bibinfo {title} {Fluctuations of the
  {B}ose–{E}instein condensate},\ }\href
  {https://doi.org/10.1088/1751-8113/47/8/085201} {\bibfield  {journal}
  {\bibinfo  {journal} {J. Phys. A Math. Theor.}\ }\textbf {\bibinfo {volume}
  {47}},\ \bibinfo {pages} {085201} (\bibinfo {year} {2014})}\BibitemShut
  {NoStop}%
\bibitem [{\citenamefont {{Smith}}(2017)}]{Smith:2017}%
  \BibitemOpen
  \bibfield  {author} {\bibinfo {author} {\bibfnamefont {R.~P.}\ \bibnamefont
  {{Smith}}},\ }\bibfield  {title} {\bibinfo {title} {Effects of interactions
  on {B}ose--{E}instein condensation},\ }in\ \href
  {https://doi.org/10.1017/9781316084366} {\emph {\bibinfo {booktitle}
  {Universal Themes of {B}ose--{E}instein Condensation}}},\ \bibinfo {editor}
  {edited by\ \bibinfo {editor} {\bibfnamefont {N.}~\bibnamefont {Proukakis}},
  \bibinfo {editor} {\bibfnamefont {D.}~\bibnamefont {Snoke}},\ and\ \bibinfo
  {editor} {\bibfnamefont {P.}~\bibnamefont {Littlewood}}}\ (\bibinfo
  {publisher} {Cambridge University Press},\ \bibinfo {year}
  {2017})\BibitemShut {NoStop}%
\bibitem [{\citenamefont {Reyes-Ayala}\ \emph {et~al.}(2019)\citenamefont
  {Reyes-Ayala}, \citenamefont {Poveda-Cuevas},\ and\ \citenamefont
  {Romero-Rochín}}]{Reyes-Ayala:2019}%
  \BibitemOpen
  \bibfield  {author} {\bibinfo {author} {\bibfnamefont {I.}~\bibnamefont
  {Reyes-Ayala}}, \bibinfo {author} {\bibfnamefont {F.~J.}\ \bibnamefont
  {Poveda-Cuevas}},\ and\ \bibinfo {author} {\bibfnamefont {V.}~\bibnamefont
  {Romero-Rochín}},\ }\bibfield  {title} {\bibinfo {title} {Non-classical
  critical exponents at {B}ose–{E}instein condensation},\ }\href
  {https://doi.org/10.1088/1742-5468/ab4984} {\bibfield  {journal} {\bibinfo
  {journal} {J. Stat. Mech. Theor. Exp.}\ }\textbf {\bibinfo {volume} {2019}},\
  \bibinfo {pages} {113102} (\bibinfo {year} {2019})}\BibitemShut {NoStop}%
\bibitem [{\citenamefont {Hadzibabic}\ and\ \citenamefont
  {Dalibard}(2011)}]{Hadzibabic:2011}%
  \BibitemOpen
  \bibfield  {author} {\bibinfo {author} {\bibfnamefont {Z.}~\bibnamefont
  {Hadzibabic}}\ and\ \bibinfo {author} {\bibfnamefont {J.}~\bibnamefont
  {Dalibard}},\ }\bibfield  {title} {\bibinfo {title} {Two-dimensional {B}ose
  fluids: An atomic physics perspective},\ }\href
  {https://doi.org/10.1393/ncr/i2011-10066-3} {\bibfield  {journal} {\bibinfo
  {journal} {Riv. Nuovo Cimento}\ }\textbf {\bibinfo {volume} {34}},\ \bibinfo
  {pages} {389} (\bibinfo {year} {2011})}\BibitemShut {NoStop}%
\bibitem [{\citenamefont {Hadzibabic}\ \emph {et~al.}(2006)\citenamefont
  {Hadzibabic}, \citenamefont {Kr{{\"u}}ger}, \citenamefont {Cheneau},
  \citenamefont {Battelier},\ and\ \citenamefont {Dalibard}}]{Hadzibabic:2006}%
  \BibitemOpen
  \bibfield  {author} {\bibinfo {author} {\bibfnamefont {Z.}~\bibnamefont
  {Hadzibabic}}, \bibinfo {author} {\bibfnamefont {P.}~\bibnamefont
  {Kr{{\"u}}ger}}, \bibinfo {author} {\bibfnamefont {M.}~\bibnamefont
  {Cheneau}}, \bibinfo {author} {\bibfnamefont {B.}~\bibnamefont {Battelier}},\
  and\ \bibinfo {author} {\bibfnamefont {J.}~\bibnamefont {Dalibard}},\
  }\bibfield  {title} {\bibinfo {title}
  {{B}erezinskii--{K}osterlitz--{T}houless crossover in a trapped atomic gas},\
  }\href {https://doi.org/10.1038/nature04851} {\bibfield  {journal} {\bibinfo
  {journal} {Nature}\ }\textbf {\bibinfo {volume} {441}},\ \bibinfo {pages}
  {1118} (\bibinfo {year} {2006})}\BibitemShut {NoStop}%
\bibitem [{\citenamefont {Jelić}\ and\ \citenamefont
  {Cugliandolo}(2011)}]{Jelic:2011}%
  \BibitemOpen
  \bibfield  {author} {\bibinfo {author} {\bibfnamefont {A.}~\bibnamefont
  {Jelić}}\ and\ \bibinfo {author} {\bibfnamefont {L.~F.}\ \bibnamefont
  {Cugliandolo}},\ }\bibfield  {title} {\bibinfo {title} {Quench dynamics of
  the 2d xy model},\ }\href {https://doi.org/10.1088/1742-5468/2011/02/P02032}
  {\bibfield  {journal} {\bibinfo  {journal} {J. Stat. Mech. Theory Exp.}\
  }\textbf {\bibinfo {volume} {2011}},\ \bibinfo {pages} {P02032} (\bibinfo
  {year} {2011})}\BibitemShut {NoStop}%
\bibitem [{\citenamefont {Fletcher}\ \emph {et~al.}(2015)\citenamefont
  {Fletcher}, \citenamefont {Robert{-}de{-}Saint{-}Vincent}, \citenamefont
  {Man}, \citenamefont {Navon}, \citenamefont {Smith}, \citenamefont
  {Viebahn},\ and\ \citenamefont {Hadzibabic}}]{Fletcher:2015}%
  \BibitemOpen
  \bibfield  {author} {\bibinfo {author} {\bibfnamefont {R.~J.}\ \bibnamefont
  {Fletcher}}, \bibinfo {author} {\bibfnamefont {M.}~\bibnamefont
  {Robert{-}de{-}Saint{-}Vincent}}, \bibinfo {author} {\bibfnamefont
  {J.}~\bibnamefont {Man}}, \bibinfo {author} {\bibfnamefont {N.}~\bibnamefont
  {Navon}}, \bibinfo {author} {\bibfnamefont {R.~P.}\ \bibnamefont {Smith}},
  \bibinfo {author} {\bibfnamefont {K.~G.~H.}\ \bibnamefont {Viebahn}},\ and\
  \bibinfo {author} {\bibfnamefont {Z.}~\bibnamefont {Hadzibabic}},\ }\bibfield
   {title} {\bibinfo {title} {Connecting {B}erezinskii-{K}osterlitz-{T}houless
  and {BEC} phase transitions by tuning interactions in a trapped gas},\ }\href
  {https://doi.org/10.1103/PhysRevLett.114.255302} {\bibfield  {journal}
  {\bibinfo  {journal} {Phys. Rev. Lett.}\ }\textbf {\bibinfo {volume} {114}},\
  \bibinfo {pages} {255302} (\bibinfo {year} {2015})}\BibitemShut {NoStop}%
\bibitem [{\citenamefont {Klaers}\ \emph
  {et~al.}(2010{\natexlab{a}})\citenamefont {Klaers}, \citenamefont {Schmitt},
  \citenamefont {Vewinger},\ and\ \citenamefont {Weitz}}]{Klaers:2010}%
  \BibitemOpen
  \bibfield  {author} {\bibinfo {author} {\bibfnamefont {J.}~\bibnamefont
  {Klaers}}, \bibinfo {author} {\bibfnamefont {J.}~\bibnamefont {Schmitt}},
  \bibinfo {author} {\bibfnamefont {F.}~\bibnamefont {Vewinger}},\ and\
  \bibinfo {author} {\bibfnamefont {M.}~\bibnamefont {Weitz}},\ }\bibfield
  {title} {\bibinfo {title} {{B}ose--{E}instein condensation of photons in an
  optical microcavity},\ }\href {https://doi.org/10.1038/nature09567}
  {\bibfield  {journal} {\bibinfo  {journal} {Nature}\ }\textbf {\bibinfo
  {volume} {468}},\ \bibinfo {pages} {545} (\bibinfo {year}
  {2010}{\natexlab{a}})}\BibitemShut {NoStop}%
\bibitem [{\citenamefont {Marelic}\ and\ \citenamefont
  {Nyman}(2015)}]{Marelic:2015}%
  \BibitemOpen
  \bibfield  {author} {\bibinfo {author} {\bibfnamefont {J.}~\bibnamefont
  {Marelic}}\ and\ \bibinfo {author} {\bibfnamefont {R.~A.}\ \bibnamefont
  {Nyman}},\ }\bibfield  {title} {\bibinfo {title} {Experimental evidence for
  inhomogeneous pumping and energy-dependent effects in photon
  {B}ose-{E}instein condensation},\ }\href
  {https://doi.org/10.1103/PhysRevA.91.033813} {\bibfield  {journal} {\bibinfo
  {journal} {Phys. Rev. A}\ }\textbf {\bibinfo {volume} {91}},\ \bibinfo
  {pages} {033813} (\bibinfo {year} {2015})}\BibitemShut {NoStop}%
\bibitem [{\citenamefont {Greveling}\ \emph {et~al.}(2018)\citenamefont
  {Greveling}, \citenamefont {Perrier},\ and\ \citenamefont {van
  Oosten}}]{Greveling:2018}%
  \BibitemOpen
  \bibfield  {author} {\bibinfo {author} {\bibfnamefont {S.}~\bibnamefont
  {Greveling}}, \bibinfo {author} {\bibfnamefont {K.~L.}\ \bibnamefont
  {Perrier}},\ and\ \bibinfo {author} {\bibfnamefont {D.}~\bibnamefont {van
  Oosten}},\ }\bibfield  {title} {\bibinfo {title} {Density distribution of a
  {B}ose-{E}instein condensate of photons in a dye-filled microcavity},\ }\href
  {https://doi.org/10.1103/PhysRevA.98.013810} {\bibfield  {journal} {\bibinfo
  {journal} {Phys. Rev. A}\ }\textbf {\bibinfo {volume} {98}},\ \bibinfo
  {pages} {013810} (\bibinfo {year} {2018})}\BibitemShut {NoStop}%
\bibitem [{\citenamefont {Balili}\ \emph {et~al.}(2007)\citenamefont {Balili},
  \citenamefont {Hartwell}, \citenamefont {Snoke}, \citenamefont {Pfeiffer},\
  and\ \citenamefont {West}}]{Balili:2007}%
  \BibitemOpen
  \bibfield  {author} {\bibinfo {author} {\bibfnamefont {R.}~\bibnamefont
  {Balili}}, \bibinfo {author} {\bibfnamefont {V.}~\bibnamefont {Hartwell}},
  \bibinfo {author} {\bibfnamefont {D.}~\bibnamefont {Snoke}}, \bibinfo
  {author} {\bibfnamefont {L.}~\bibnamefont {Pfeiffer}},\ and\ \bibinfo
  {author} {\bibfnamefont {K.}~\bibnamefont {West}},\ }\bibfield  {title}
  {\bibinfo {title} {Bose-einstein condensation of microcavity polaritons in a
  trap},\ }\href {https://doi.org/10.1126/science.1140990} {\bibfield
  {journal} {\bibinfo  {journal} {Science}\ }\textbf {\bibinfo {volume}
  {316}},\ \bibinfo {pages} {1007} (\bibinfo {year} {2007})}\BibitemShut
  {NoStop}%
\bibitem [{\citenamefont {Gaunt}\ \emph {et~al.}(2013)\citenamefont {Gaunt},
  \citenamefont {Schmidutz}, \citenamefont {Gotlibovych}, \citenamefont
  {Smith},\ and\ \citenamefont {Hadzibabic}}]{Gaunt:2013}%
  \BibitemOpen
  \bibfield  {author} {\bibinfo {author} {\bibfnamefont {A.~L.}\ \bibnamefont
  {Gaunt}}, \bibinfo {author} {\bibfnamefont {T.~F.}\ \bibnamefont
  {Schmidutz}}, \bibinfo {author} {\bibfnamefont {I.}~\bibnamefont
  {Gotlibovych}}, \bibinfo {author} {\bibfnamefont {R.~P.}\ \bibnamefont
  {Smith}},\ and\ \bibinfo {author} {\bibfnamefont {Z.}~\bibnamefont
  {Hadzibabic}},\ }\bibfield  {title} {\bibinfo {title} {{{B}ose--{E}instein
  Condensation of Atoms in a Uniform Potential}},\ }\href
  {https://doi.org/10.1103/PhysRevLett.110.200406} {\bibfield  {journal}
  {\bibinfo  {journal} {Phys. Rev. Lett.}\ }\textbf {\bibinfo {volume} {110}},\
  \bibinfo {pages} {200406} (\bibinfo {year} {2013})}\BibitemShut {NoStop}%
\bibitem [{\citenamefont {Busley}\ \emph {et~al.}(2022)\citenamefont {Busley},
  \citenamefont {{Espert Miranda}}, \citenamefont {Redmann}, \citenamefont
  {Kurtscheid}, \citenamefont {{Karkihalli Umesh}}, \citenamefont {Vewinger},
  \citenamefont {Weitz},\ and\ \citenamefont {Schmitt}}]{Busley:2022}%
  \BibitemOpen
  \bibfield  {author} {\bibinfo {author} {\bibfnamefont {E.}~\bibnamefont
  {Busley}}, \bibinfo {author} {\bibfnamefont {L.}~\bibnamefont {{Espert
  Miranda}}}, \bibinfo {author} {\bibfnamefont {A.}~\bibnamefont {Redmann}},
  \bibinfo {author} {\bibfnamefont {C.}~\bibnamefont {Kurtscheid}}, \bibinfo
  {author} {\bibfnamefont {K.}~\bibnamefont {{Karkihalli Umesh}}}, \bibinfo
  {author} {\bibfnamefont {F.}~\bibnamefont {Vewinger}}, \bibinfo {author}
  {\bibfnamefont {M.}~\bibnamefont {Weitz}},\ and\ \bibinfo {author}
  {\bibfnamefont {J.}~\bibnamefont {Schmitt}},\ }\bibfield  {title} {\bibinfo
  {title} {Compressibility and the equation of state of an optical quantum gas
  in a box},\ }\href {https://doi.org/10.1126/science.abm2543} {\bibfield
  {journal} {\bibinfo  {journal} {Science}\ }\textbf {\bibinfo {volume}
  {375}},\ \bibinfo {pages} {1403} (\bibinfo {year} {2022})}\BibitemShut
  {NoStop}%
\bibitem [{\citenamefont {Comaron}\ \emph {et~al.}(2025)\citenamefont
  {Comaron}, \citenamefont {Estrecho}, \citenamefont {Wurdack}, \citenamefont
  {Pieczarka}, \citenamefont {Steger}, \citenamefont {Snoke}, \citenamefont
  {West}, \citenamefont {Pfeiffer}, \citenamefont {Truscott}, \citenamefont
  {Matuszewski}, \citenamefont {Szyma{\'n}ska},\ and\ \citenamefont
  {Ostrovskaya}}]{Comaron:2025}%
  \BibitemOpen
  \bibfield  {author} {\bibinfo {author} {\bibfnamefont {P.}~\bibnamefont
  {Comaron}}, \bibinfo {author} {\bibfnamefont {E.}~\bibnamefont {Estrecho}},
  \bibinfo {author} {\bibfnamefont {M.}~\bibnamefont {Wurdack}}, \bibinfo
  {author} {\bibfnamefont {M.}~\bibnamefont {Pieczarka}}, \bibinfo {author}
  {\bibfnamefont {M.}~\bibnamefont {Steger}}, \bibinfo {author} {\bibfnamefont
  {D.~W.}\ \bibnamefont {Snoke}}, \bibinfo {author} {\bibfnamefont
  {K.}~\bibnamefont {West}}, \bibinfo {author} {\bibfnamefont {L.~N.}\
  \bibnamefont {Pfeiffer}}, \bibinfo {author} {\bibfnamefont {A.~G.}\
  \bibnamefont {Truscott}}, \bibinfo {author} {\bibfnamefont {M.}~\bibnamefont
  {Matuszewski}}, \bibinfo {author} {\bibfnamefont {M.~H.}\ \bibnamefont
  {Szyma{\'n}ska}},\ and\ \bibinfo {author} {\bibfnamefont {E.~A.}\
  \bibnamefont {Ostrovskaya}},\ }\bibfield  {title} {\bibinfo {title}
  {Coherence of a non-equilibrium polariton condensate across the
  interaction-mediated phase transition},\ }\href
  {https://doi.org/10.1038/s42005-025-01977-7} {\bibfield  {journal} {\bibinfo
  {journal} {Communications Physics}\ }\textbf {\bibinfo {volume} {8}},\
  \bibinfo {pages} {94} (\bibinfo {year} {2025})}\BibitemShut {NoStop}%
\bibitem [{\citenamefont {Kurtscheid}\ \emph {et~al.}(2020)\citenamefont
  {Kurtscheid}, \citenamefont {Dung}, \citenamefont {Redmann}, \citenamefont
  {Busley}, \citenamefont {Klaers}, \citenamefont {Vewinger}, \citenamefont
  {Schmitt},\ and\ \citenamefont {Weitz}}]{Kurtscheid:2020}%
  \BibitemOpen
  \bibfield  {author} {\bibinfo {author} {\bibfnamefont {C.}~\bibnamefont
  {Kurtscheid}}, \bibinfo {author} {\bibfnamefont {D.}~\bibnamefont {Dung}},
  \bibinfo {author} {\bibfnamefont {A.}~\bibnamefont {Redmann}}, \bibinfo
  {author} {\bibfnamefont {E.}~\bibnamefont {Busley}}, \bibinfo {author}
  {\bibfnamefont {J.}~\bibnamefont {Klaers}}, \bibinfo {author} {\bibfnamefont
  {F.}~\bibnamefont {Vewinger}}, \bibinfo {author} {\bibfnamefont
  {J.}~\bibnamefont {Schmitt}},\ and\ \bibinfo {author} {\bibfnamefont
  {M.}~\bibnamefont {Weitz}},\ }\bibfield  {title} {\bibinfo {title} {Realizing
  arbitrary trapping potentials for light via direct laser writing of mirror
  surface profiles},\ }\href {https://doi.org/10.1209/0295-5075/130/54001}
  {\bibfield  {journal} {\bibinfo  {journal} {{EPL} (Europhysics Letters)}\
  }\textbf {\bibinfo {volume} {130}},\ \bibinfo {pages} {54001} (\bibinfo
  {year} {2020})}\BibitemShut {NoStop}%
\bibitem [{\citenamefont {Vretenar}\ \emph {et~al.}(2023)\citenamefont
  {Vretenar}, \citenamefont {Puplauskis},\ and\ \citenamefont
  {Klaers}}]{Vretenar:2023}%
  \BibitemOpen
  \bibfield  {author} {\bibinfo {author} {\bibfnamefont {M.}~\bibnamefont
  {Vretenar}}, \bibinfo {author} {\bibfnamefont {M.}~\bibnamefont
  {Puplauskis}},\ and\ \bibinfo {author} {\bibfnamefont {J.}~\bibnamefont
  {Klaers}},\ }\bibfield  {title} {\bibinfo {title} {Mirror surface
  nanostructuring via laser direct writing—characterization and physical
  origins},\ }\href {https://doi.org/https://doi.org/10.1002/adom.202202820}
  {\bibfield  {journal} {\bibinfo  {journal} {Adv. Opt. Mater.}\ }\textbf
  {\bibinfo {volume} {11}},\ \bibinfo {pages} {2202820} (\bibinfo {year}
  {2023})}\BibitemShut {NoStop}%
\bibitem [{\citenamefont {Schmitt}\ \emph {et~al.}(2015)\citenamefont
  {Schmitt}, \citenamefont {Damm}, \citenamefont {Dung}, \citenamefont
  {Vewinger}, \citenamefont {Klaers},\ and\ \citenamefont
  {Weitz}}]{Schmitt:2015}%
  \BibitemOpen
  \bibfield  {author} {\bibinfo {author} {\bibfnamefont {J.}~\bibnamefont
  {Schmitt}}, \bibinfo {author} {\bibfnamefont {T.}~\bibnamefont {Damm}},
  \bibinfo {author} {\bibfnamefont {D.}~\bibnamefont {Dung}}, \bibinfo {author}
  {\bibfnamefont {F.}~\bibnamefont {Vewinger}}, \bibinfo {author}
  {\bibfnamefont {J.}~\bibnamefont {Klaers}},\ and\ \bibinfo {author}
  {\bibfnamefont {M.}~\bibnamefont {Weitz}},\ }\bibfield  {title} {\bibinfo
  {title} {Thermalization kinetics of light: From laser dynamics to equilibrium
  condensation of photons},\ }\href
  {https://doi.org/10.1103/PhysRevA.92.011602} {\bibfield  {journal} {\bibinfo
  {journal} {Phys. Rev. A}\ }\textbf {\bibinfo {volume} {92}},\ \bibinfo
  {pages} {011602} (\bibinfo {year} {2015})}\BibitemShut {NoStop}%
\bibitem [{\citenamefont {Schmitt}(2018)}]{Schmitt:2018}%
  \BibitemOpen
  \bibfield  {author} {\bibinfo {author} {\bibfnamefont {J.}~\bibnamefont
  {Schmitt}},\ }\bibfield  {title} {\bibinfo {title} {Dynamics and correlations
  of a {B}ose{\textendash}{E}instein condensate of photons},\ }\href
  {https://doi.org/10.1088/1361-6455/aad409} {\bibfield  {journal} {\bibinfo
  {journal} {J. Phys. B: At. Mol. Opt. Phys.}\ }\textbf {\bibinfo {volume}
  {51}},\ \bibinfo {pages} {173001} (\bibinfo {year} {2018})}\BibitemShut
  {NoStop}%
\bibitem [{\citenamefont {Kirton}\ and\ \citenamefont
  {Keeling}(2013)}]{Kirton:2013}%
  \BibitemOpen
  \bibfield  {author} {\bibinfo {author} {\bibfnamefont {P.}~\bibnamefont
  {Kirton}}\ and\ \bibinfo {author} {\bibfnamefont {J.}~\bibnamefont
  {Keeling}},\ }\bibfield  {title} {\bibinfo {title} {Nonequilibrium model of
  photon condensation},\ }\href
  {https://doi.org/10.1103/PhysRevLett.111.100404} {\bibfield  {journal}
  {\bibinfo  {journal} {Phys. Rev. Lett.}\ }\textbf {\bibinfo {volume} {111}},\
  \bibinfo {pages} {100404} (\bibinfo {year} {2013})}\BibitemShut {NoStop}%
\bibitem [{\citenamefont {Dung}\ \emph {et~al.}(2017)\citenamefont {Dung},
  \citenamefont {Kurtscheid}, \citenamefont {Damm}, \citenamefont {Schmitt},
  \citenamefont {Vewinger}, \citenamefont {Weitz},\ and\ \citenamefont
  {Klaers}}]{Dung:2017}%
  \BibitemOpen
  \bibfield  {author} {\bibinfo {author} {\bibfnamefont {D.}~\bibnamefont
  {Dung}}, \bibinfo {author} {\bibfnamefont {C.}~\bibnamefont {Kurtscheid}},
  \bibinfo {author} {\bibfnamefont {T.}~\bibnamefont {Damm}}, \bibinfo {author}
  {\bibfnamefont {J.}~\bibnamefont {Schmitt}}, \bibinfo {author} {\bibfnamefont
  {F.}~\bibnamefont {Vewinger}}, \bibinfo {author} {\bibfnamefont
  {M.}~\bibnamefont {Weitz}},\ and\ \bibinfo {author} {\bibfnamefont
  {J.}~\bibnamefont {Klaers}},\ }\bibfield  {title} {\bibinfo {title} {Variable
  potentials for thermalized light and coupled condensates},\ }\href
  {https://doi.org/10.1038/nphoton.2017.139} {\bibfield  {journal} {\bibinfo
  {journal} {Nat. Photon.}\ }\textbf {\bibinfo {volume} {11}},\ \bibinfo
  {pages} {565} (\bibinfo {year} {2017})}\BibitemShut {NoStop}%
\bibitem [{\citenamefont {Trypogeorgos}\ \emph {et~al.}(2025)\citenamefont
  {Trypogeorgos}, \citenamefont {Gianfrate}, \citenamefont {Landini},
  \citenamefont {Nigro}, \citenamefont {Gerace}, \citenamefont {Carusotto},
  \citenamefont {Riminucci}, \citenamefont {Baldwin}, \citenamefont {Pfeiffer},
  \citenamefont {Martone}, \citenamefont {De~Giorgi}, \citenamefont
  {Ballarini},\ and\ \citenamefont {Sanvitto}}]{Trypogeorgos:2025}%
  \BibitemOpen
  \bibfield  {author} {\bibinfo {author} {\bibfnamefont {D.}~\bibnamefont
  {Trypogeorgos}}, \bibinfo {author} {\bibfnamefont {A.}~\bibnamefont
  {Gianfrate}}, \bibinfo {author} {\bibfnamefont {M.}~\bibnamefont {Landini}},
  \bibinfo {author} {\bibfnamefont {D.}~\bibnamefont {Nigro}}, \bibinfo
  {author} {\bibfnamefont {D.}~\bibnamefont {Gerace}}, \bibinfo {author}
  {\bibfnamefont {I.}~\bibnamefont {Carusotto}}, \bibinfo {author}
  {\bibfnamefont {F.}~\bibnamefont {Riminucci}}, \bibinfo {author}
  {\bibfnamefont {K.~W.}\ \bibnamefont {Baldwin}}, \bibinfo {author}
  {\bibfnamefont {L.~N.}\ \bibnamefont {Pfeiffer}}, \bibinfo {author}
  {\bibfnamefont {G.~I.}\ \bibnamefont {Martone}}, \bibinfo {author}
  {\bibfnamefont {M.}~\bibnamefont {De~Giorgi}}, \bibinfo {author}
  {\bibfnamefont {D.}~\bibnamefont {Ballarini}},\ and\ \bibinfo {author}
  {\bibfnamefont {D.}~\bibnamefont {Sanvitto}},\ }\bibfield  {title} {\bibinfo
  {title} {Emerging supersolidity in photonic-crystal polariton condensates},\
  }\href {https://doi.org/10.1038/s41586-025-08616-9} {\bibfield  {journal}
  {\bibinfo  {journal} {Nature}\ }\textbf {\bibinfo {volume} {639}},\ \bibinfo
  {pages} {337} (\bibinfo {year} {2025})}\BibitemShut {NoStop}%
\bibitem [{\citenamefont {P{\"o}schl}\ and\ \citenamefont
  {Teller}(1933)}]{Poeschl:1933}%
  \BibitemOpen
  \bibfield  {author} {\bibinfo {author} {\bibfnamefont {G.}~\bibnamefont
  {P{\"o}schl}}\ and\ \bibinfo {author} {\bibfnamefont {E.}~\bibnamefont
  {Teller}},\ }\bibfield  {title} {\bibinfo {title} {{B}emerkungen zur
  {Q}uantenmechanik des anharmonischen {O}szillators},\ }\href
  {https://doi.org/10.1007/BF01331132} {\bibfield  {journal} {\bibinfo
  {journal} {Z. Phys.}\ }\textbf {\bibinfo {volume} {83}},\ \bibinfo {pages}
  {143} (\bibinfo {year} {1933})}\BibitemShut {NoStop}%
\bibitem [{\citenamefont {Nieto}(1978)}]{Nieto:1978}%
  \BibitemOpen
  \bibfield  {author} {\bibinfo {author} {\bibfnamefont {M.~M.}\ \bibnamefont
  {Nieto}},\ }\bibfield  {title} {\bibinfo {title} {Exact wave-function
  normalization constants for the {${B}_{0}\tanh
  z\ensuremath{-}{U}_{0}{\cosh}^{\ensuremath{-}2} z$} and {P}\"oschl-{T}eller
  potentials},\ }\href {https://doi.org/10.1103/PhysRevA.17.1273} {\bibfield
  {journal} {\bibinfo  {journal} {Phys. Rev. A}\ }\textbf {\bibinfo {volume}
  {17}},\ \bibinfo {pages} {1273} (\bibinfo {year} {1978})}\BibitemShut
  {NoStop}%
\bibitem [{\citenamefont {Wolf}\ and\ \citenamefont
  {Carter}(1975)}]{Wolf:1975}%
  \BibitemOpen
  \bibfield  {author} {\bibinfo {author} {\bibfnamefont {E.}~\bibnamefont
  {Wolf}}\ and\ \bibinfo {author} {\bibfnamefont {W.~H.}\ \bibnamefont
  {Carter}},\ }\bibfield  {title} {\bibinfo {title} {Angular distribution of
  radiant intensity from sources of different degrees of spatial coherence},\
  }\href {https://doi.org/https://doi.org/10.1016/0030-4018(75)90081-4}
  {\bibfield  {journal} {\bibinfo  {journal} {Opt. Commun.}\ }\textbf {\bibinfo
  {volume} {13}},\ \bibinfo {pages} {205} (\bibinfo {year} {1975})}\BibitemShut
  {NoStop}%
\bibitem [{\citenamefont {Tammuz}\ \emph {et~al.}(2011)\citenamefont {Tammuz},
  \citenamefont {Smith}, \citenamefont {Campbell}, \citenamefont {Beattie},
  \citenamefont {Moulder}, \citenamefont {Dalibard},\ and\ \citenamefont
  {Hadzibabic}}]{Tammuz:2011}%
  \BibitemOpen
  \bibfield  {author} {\bibinfo {author} {\bibfnamefont {N.}~\bibnamefont
  {Tammuz}}, \bibinfo {author} {\bibfnamefont {R.~P.}\ \bibnamefont {Smith}},
  \bibinfo {author} {\bibfnamefont {R.~L.~D.}\ \bibnamefont {Campbell}},
  \bibinfo {author} {\bibfnamefont {S.}~\bibnamefont {Beattie}}, \bibinfo
  {author} {\bibfnamefont {S.}~\bibnamefont {Moulder}}, \bibinfo {author}
  {\bibfnamefont {J.}~\bibnamefont {Dalibard}},\ and\ \bibinfo {author}
  {\bibfnamefont {Z.}~\bibnamefont {Hadzibabic}},\ }\bibfield  {title}
  {\bibinfo {title} {Can a {B}ose gas be saturated?},\ }\href
  {https://doi.org/10.1103/PhysRevLett.106.230401} {\bibfield  {journal}
  {\bibinfo  {journal} {Phys. Rev. Lett.}\ }\textbf {\bibinfo {volume} {106}},\
  \bibinfo {pages} {230401} (\bibinfo {year} {2011})}\BibitemShut {NoStop}%
\bibitem [{\citenamefont {Hanai}\ and\ \citenamefont
  {Littlewood}(2020)}]{Hanai:2020}%
  \BibitemOpen
  \bibfield  {author} {\bibinfo {author} {\bibfnamefont {R.}~\bibnamefont
  {Hanai}}\ and\ \bibinfo {author} {\bibfnamefont {P.~B.}\ \bibnamefont
  {Littlewood}},\ }\bibfield  {title} {\bibinfo {title} {Critical fluctuations
  at a many-body exceptional point},\ }\href
  {https://doi.org/10.1103/PhysRevResearch.2.033018} {\bibfield  {journal}
  {\bibinfo  {journal} {Phys. Rev. Res.}\ }\textbf {\bibinfo {volume} {2}},\
  \bibinfo {pages} {033018} (\bibinfo {year} {2020})}\BibitemShut {NoStop}%
\bibitem [{\citenamefont {Zelle}\ \emph {et~al.}(2024)\citenamefont {Zelle},
  \citenamefont {Daviet}, \citenamefont {Rosch},\ and\ \citenamefont
  {Diehl}}]{Zelle:2024}%
  \BibitemOpen
  \bibfield  {author} {\bibinfo {author} {\bibfnamefont {C.~P.}\ \bibnamefont
  {Zelle}}, \bibinfo {author} {\bibfnamefont {R.}~\bibnamefont {Daviet}},
  \bibinfo {author} {\bibfnamefont {A.}~\bibnamefont {Rosch}},\ and\ \bibinfo
  {author} {\bibfnamefont {S.}~\bibnamefont {Diehl}},\ }\bibfield  {title}
  {\bibinfo {title} {Universal phenomenology at critical exceptional points of
  nonequilibrium $\mathrm{O}(n)$ models},\ }\href
  {https://doi.org/10.1103/PhysRevX.14.021052} {\bibfield  {journal} {\bibinfo
  {journal} {Phys. Rev. X}\ }\textbf {\bibinfo {volume} {14}},\ \bibinfo
  {pages} {021052} (\bibinfo {year} {2024})}\BibitemShut {NoStop}%
\bibitem [{\citenamefont {Berges}\ \emph {et~al.}(2015)\citenamefont {Berges},
  \citenamefont {Boguslavski}, \citenamefont {Schlichting},\ and\ \citenamefont
  {Venugopalan}}]{Berges:2015b}%
  \BibitemOpen
  \bibfield  {author} {\bibinfo {author} {\bibfnamefont {J.}~\bibnamefont
  {Berges}}, \bibinfo {author} {\bibfnamefont {K.}~\bibnamefont {Boguslavski}},
  \bibinfo {author} {\bibfnamefont {S.}~\bibnamefont {Schlichting}},\ and\
  \bibinfo {author} {\bibfnamefont {R.}~\bibnamefont {Venugopalan}},\
  }\bibfield  {title} {\bibinfo {title} {Universality far from equilibrium:
  From superfluid {B}ose gases to heavy-ion collisions},\ }\href
  {https://doi.org/10.1103/PhysRevLett.114.061601} {\bibfield  {journal}
  {\bibinfo  {journal} {Phys. Rev. Lett.}\ }\textbf {\bibinfo {volume} {114}},\
  \bibinfo {pages} {061601} (\bibinfo {year} {2015})}\BibitemShut {NoStop}%
\bibitem [{\citenamefont {Klaers}\ \emph
  {et~al.}(2010{\natexlab{b}})\citenamefont {Klaers}, \citenamefont
  {Vewinger},\ and\ \citenamefont {Weitz}}]{Klaers:2010b}%
  \BibitemOpen
  \bibfield  {author} {\bibinfo {author} {\bibfnamefont {J.}~\bibnamefont
  {Klaers}}, \bibinfo {author} {\bibfnamefont {F.}~\bibnamefont {Vewinger}},\
  and\ \bibinfo {author} {\bibfnamefont {M.}~\bibnamefont {Weitz}},\ }\bibfield
   {title} {\bibinfo {title} {Thermalization of a two-dimensional photonic gas
  in a `white wall' photon box},\ }\href {https://doi.org/10.1038/nphys1680}
  {\bibfield  {journal} {\bibinfo  {journal} {Nature Physics}\ }\textbf
  {\bibinfo {volume} {6}},\ \bibinfo {pages} {512} (\bibinfo {year}
  {2010}{\natexlab{b}})}\BibitemShut {NoStop}%
\bibitem [{\citenamefont {Naraschewski}\ and\ \citenamefont
  {Glauber}(1999)}]{Naraschewski:1999}%
  \BibitemOpen
  \bibfield  {author} {\bibinfo {author} {\bibfnamefont {M.}~\bibnamefont
  {Naraschewski}}\ and\ \bibinfo {author} {\bibfnamefont {R.~J.}\ \bibnamefont
  {Glauber}},\ }\bibfield  {title} {\bibinfo {title} {Spatial coherence and
  density correlations of trapped {B}ose gases},\ }\href
  {https://doi.org/10.1103/PhysRevA.59.4595} {\bibfield  {journal} {\bibinfo
  {journal} {Phys. Rev. A}\ }\textbf {\bibinfo {volume} {59}},\ \bibinfo
  {pages} {4595} (\bibinfo {year} {1999})}\BibitemShut {NoStop}%
\bibitem [{\citenamefont {Morales-Amador}\ \emph {et~al.}(2024)\citenamefont
  {Morales-Amador}, \citenamefont {Romero-Rochín},\ and\ \citenamefont
  {Paredes}}]{Morales-Amador:2024}%
  \BibitemOpen
  \bibfield  {author} {\bibinfo {author} {\bibfnamefont {M.~I.}\ \bibnamefont
  {Morales-Amador}}, \bibinfo {author} {\bibfnamefont {V.}~\bibnamefont
  {Romero-Rochín}},\ and\ \bibinfo {author} {\bibfnamefont {R.}~\bibnamefont
  {Paredes}},\ }\bibfield  {title} {\bibinfo {title} {Critical exponents and
  fluctuations at {BEC} in a 2{D} harmonically trapped ideal gas},\ }\href
  {https://doi.org/10.1088/1361-6455/ad2860} {\bibfield  {journal} {\bibinfo
  {journal} {J. Phys. B: At. Mol. Opt. Phys.}\ }\textbf {\bibinfo {volume}
  {57}},\ \bibinfo {pages} {045301} (\bibinfo {year} {2024})}\BibitemShut
  {NoStop}%
\end{thebibliography}%

\vspace{0.3cm}
\noindent \textit{Acknowledgments---} We thank Nir Davidson, James Anglin, Jean Dalibard, Thierry Giamarchi, Corinna Kollath and Carl Zelle for fruitful discussions, and Maciej Ga\l ka, Achim Rosch and Daniele Sanvitto for comments on the manuscript. The authors acknowledge funding from the DFG within SFB/TR 185 (277625399), Cluster of Excellence ML4Q (EXC 2004/1–390534769) and STRUCTURES (EXC 2181, 390900948), from the EU (ERC, TopoGrand, 101040409), and from the grants UNAM-PAPIIT IN-117623 and IA-101325.

\noindent \textit{Data availability---} All data needed to evaluate and reproduce the conclusions in the paper are present in the paper and/or the SI. The data that support the findings of this study are available in the Zenodo repository (\href{https://doi.org/10.5281/zenodo.20097108}{10.5281/zenodo.20097109}).

\setcounter{figure}{0} 
\setcounter{equation}{0} 
\renewcommand\theequation{S\arabic{equation}} 
\renewcommand\thefigure{S\arabic{figure}} 

\vspace{0cm}
\section*{Supplementary Information}
\vspace{-0.1cm}

\subsection{1. Experimental scheme}
\vspace{-0.1cm}

The two-dimensional photon gas in the box potential is created within a high-finesse optical microcavity, consisting of a plane and a surface-structured mirror with reflectivity $>99.9985\%$ spaced by $D_0\approx \SI{1.6}{\micro\meter}$. The cavity is filled with a dye solution (Rhodamine 6G in ethylene glycol, concentration between 0.03 and $\SI{3}{\milli\mol\per\liter}$, refractive index $\tilde n=1.44$) which fulfils the Kennard-Stepanov relation and allows for a photon thermalization at $T = \SI{300}{\kelvin}$ after repeated absorption and emission processes~\cite{Klaers:2010,Klaers:2010b}. At a large dye concentration of $\SI{1}{\milli\mol\per\liter}$, the cavity lifetime $\tau_\mathrm{cav}\approx \SI{0.35}{\nano\second}$ exceeds the thermalization time $\tau_\mathrm{cav}\approx\SI{0.11}{\nano\second}$, so that the system is well equilibrated and exhibits a thermal occupation of the excited $k$ states shown in Fig.~\ref{fig:2}(B)~\cite{Schmitt:2015,Schmitt:2018}. To tune the system out of equilibrium [see Fig.~\ref{fig:4}(C) of the main text], the dye concentration was changed to $\{3, 0.3, 0.1, 0.03\}\SI{}{\milli\mol\per\liter}$, thereby altering the thermalization time $\tau_\mathrm{th}\approx \{0.04, 0.3, 1, 3\}\SI{}{\nano\second}$. The free spectral range of the microcavity of $\SI{65}{\tera\hertz}$ exceeds thermal energy $\kB T\approx h\times \SI{6.2}{\tera\hertz}$ (in frequency units), such that only photons from one longitudinal mode interact with the dye, freezing out the longitudinal degree of freedom. The photon dispersion becomes two-dimensional,
\begin{equation}
    E\approx \frac{\hbar^2(k_x^2+k_y^2)}{2m}+ m \left(\frac{c}{\tilde n}\right)^2\left[\frac{h(x,y)}{D_0}\right]
    \label{eq:S1}
\end{equation}
with an effective photon mass $m$, refractive index $\tilde n$ and surface height profile $h(x,y)$; the latter is imprinted on one of the cavity mirrors and realizes the potential. To imprint surface elevations which act as a repulsive potential, the mirror coating contains a $\SI{30}{\nano\meter}$-thin Si layer, which can be heated in a spatially resolved way by scanning a $\SI{532}{\nano\meter}$-wavelength laser beam across the mirror sample. The heating causes the dielectric coating to permanently lift without notably affecting the mirror reflection properties~\cite{Kurtscheid:2020,Vretenar:2023}. Using this method, we realize different-sized square-shaped box potentials for photons with side lengths $L = \{40, 50, 60, 80\}\SI{}{\micro\meter}$~\cite{Busley:2022}. The potential depth $\UD = 1.4 \kB T$ of the potential is given by the $\SI{27}{\nano\meter}$ surface height, and the $10$-$90\%$ width of the walls is approximately $\SI{3}{\micro\meter}$. The experimental surface profiles of the imprinted structures on the cavity mirrors at low elevation heights can be well described by a Pöschl-Teller potential with a value of $\alpha\approx 1.5$, as discussed in the main text and Sections 4 and 5 of the SI, corresponding to a box trap with slightly rounded edges near the bottom [see. Fig.~\ref{fig:1}(B) of the main text].

\vspace{-0.2cm}
\subsection{2. First-order spatial correlations}
\vspace{-0.1cm}

The spatial correlation function for the bosonic field $\Psi(\mathbf{r},t) = \sum_i c_i(t) \psi_i(\mathbf{r})$ at equal times is defined as $G^{(1)}(\mathbf{r},\mathbf{r'}) = \langle \Psi^*(\mathbf{r},t)\Psi(\mathbf{r'},t)\rangle_t$, where $\psi_i(\mathbf{r})$ are the eigenstates in the trap populated by the photons, yielding $G^{(1)}(\mathbf{r},\mathbf{r'}) = \sum_{i,j} \psi_i^*(\mathbf{r})\psi_j(\mathbf{r'})\langle c_i^*(t)c_j(t)\rangle$. Due to the frequent absorption and re-emission processes of the photons by the dye molecules, which thermalize the photon gas, cross-correlations $\langle c_i^*(t)c_j(t)\rangle = \delta_{ij}\langle|c_i(t)|^2\rangle = \delta_{ij}\bar n_i$ vanish on temporal average. Only the mean population $\bar n_i$ of the $i$-th eigenstate contributes to the correlation function
\begin{equation}    
    G^{(1)}(\mathbf{r},\mathbf{r'})=\sum_i \bar n_i \psi_i^*(\mathbf{r})\psi_i(\mathbf{r'}).
    \label{eq:S2}
\end{equation}
To quantify the global degree of coherence, one is interested in the average relative correlations between any two points $\mathbf{r}$ and $\mathbf{r'}$ in the gas, given by the volume-averaged functions~\cite{Naraschewski:1999}
\begin{widetext}
\begin{subequations}
\begin{align}
    \label{eq:S3a}
    G^{(1)}_{\textrm{va}}({\bf s}) &= \int d{\bf R}~G^{(1)}\left( {\bf R}-\frac{\bf s}{2},{\bf R}+\frac{\bf s}{2}\right),\\
    g^{(1)}_{\textrm{va}}({\bf s}) &= \frac{\int d{\bf R}~G^{(1)}\left( {\bf R}-\frac{\bf s}{2},{\bf R}+\frac{\bf s}{2}\right)}{\int d{\bf R}~\sqrt{G^{(1)}\left({\bf R}-\frac{\bf s}{2}, {\bf R}-\frac{\bf s}{2}\right)G^{(1)}\left({\bf R}+\frac{\bf s}{2},{\bf R}+\frac{\bf s}{2}\right)}},
    \label{eq:S3b}
\end{align}
\end{subequations}
\end{widetext}
where $\bf{s}={\bf r-r'}$ denotes the vector between two points centered around $\mathbf{R}$, which is integrated over the volume. Expanding the eigenfunctions in position space $\psi_i({\bf r}) = 1/(2\pi)\int \phi_i({\bf k})e^{i{\bf k}{\bf r}}d^2{\bf k}$, the correlation function in eq.~\eqref{eq:S2} becomes 
\begin{equation}
    {G}^{(1)}({\bf s}) = \frac{1}{(2\pi)^2}\int \bar{n}_k({\bf k}) e^{i{\bf k}{\bf s}}d^2{\bf k},
    \label{eq:S4}
\end{equation}
where $\bar{n}_k({\bf k})=\sum_i \bar{n}_i |\phi_i({\bf k})|^2$ is the momentum-space distribution of the 2D photon gas. We here have assumed that the momentum-space eigenfunctions $\phi_i(\mathbf{k})$ are orthogonal, which is well satisfied for both rigid box potentials, as well as for our experimentally realized box potentials. Notably, expression eq.~\eqref{eq:S4} considers all possible pairs of points in the photon gas (separated by vector $\mathbf{s}$) which contribute to the total correlation function and is therefore equivalent to eq.~\eqref{eq:S3a}. The degree of coherence $g^{(1)}({\bf s})$ immediately follows from normalization, as in eq.~\eqref{eq:S3b}. For the case of a gas with nearly uniform density $n$, the denominator trivially becomes $\int d\mathbf{R}~G^{(1)}(\mathbf{R},\mathbf{R})\approx N_\mathrm{pairs}(\mathbf{s})\cdot n$, where $N_\mathrm{pairs}(\mathbf{s})$ denotes the number of pairs of points spaced by $\mathbf{s}$ that fit into the system:
\begin{equation}
    g^{(1)}({\bf s}) = \frac{G^{(1)}({\bf s})}{N_{\textrm{pairs}}({\bf s})\cdot n}.
    \label{eq:S5}
\end{equation}
For the infinite system, one has $N_\mathrm{pairs}(\mathbf{s})=\mathrm{const.}$, because for every distance $\mathbf{s}$, the same number of pairs (i.e., infinitely many) can be found. For finite sizes, on the other hand, $N_\mathrm{pairs}(\mathbf{s})$ depends explicitly on the distance $\mathbf{s}$, each of which contributes a different statistical weight. Finally, we remark that from an experimental perspective, the Fourier-based determination of the first-order correlations described above offers important advantages over interferometric methods. A key benefit is the possibility to use a single-shot detection instead of requiring long integration times, allowing us to resolve the critical region close to the phase transition under typical experimental stability conditions.

\begin{figure}[t]
  \centering
  \includegraphics[width=1\columnwidth]{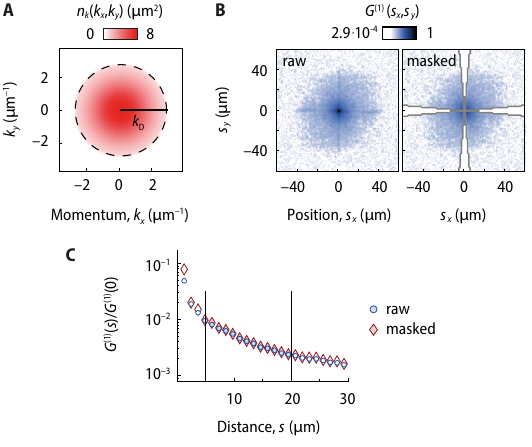}
 \caption{$\vert$ \textbf{Momentum space distribution and spatial correlations. (A)} Theoretically calculated momentum space distribution deep in the uncondensed phase for a chemical potential $\mu = -\kB T$. The distribution is circularly symmetric and shows the occupation of wave vectors with $|\mathbf{k}|<\kD$. \textbf{(B)} Raw and masked correlation data $G^{(1)}(s_x,s_y)$. The masking tests the influence of the cross-shaped features visible at $s_x=0$ and $s_y=0$, respectively, on the fitting of $\xi$. \textbf{(C)} The azimuthally-averaged correlations $G^{(1)}(s)$ are barely affected by the masking, indicating that the cross features do not obscure the extraction of $\xi$.}
\label{fig:5}
\end{figure}

\subsection{3. Extraction of the correlation length}

To inject the photon gas, the dye medium is pumped with a $\SI{532}{\nano\meter}$ laser beam, which is chopped into $\SI{500}{\nano\second}$-long pulses, more than two orders of magnitude longer than the thermalization time~\cite{Schmitt:2015}, at a $\SI{30}{\hertz}$ repetition rate. By ramping up the pump power over the course of many pulses, the total photon number $N$ in the cavity in subsequent pulses is increased until the critical value $\Nc$ is reached. This allows us to prepare the gas at different reduced temperatures $t$, especially those near the phase transition at $t = 0$. Within each pulse the photon number is approximately constant. To determine the temperature-dependent first-order correlations $G^{(1)}(\mathbf{s})$ in the uniform gas between two points separated by $\mathbf{s} = (s_x,s_y) = (x-x',y-y')$, the far-field cavity emission is imaged onto a camera positioned in the Fourier plane of a lens, where the momentum distribution $n_k(\mathbf{k})$ is recorded. To improve the signal-to-noise ratio, we average over 5 to 20 consecutive experiment cycles. The measured $n_k(\mathbf{k})$ exhibits a cutoff at $\kD \approx \SI{2.9}{\per\micro\meter}$ imposed by the finite trap depth $\UD$, see Fig.~\ref{fig:2}(A) of the main text. As this cutoff causes unphysical artefacts and we are only interested in the Lorentzian part of the momentum distribution in the quantum-degenerate regime ($n\lambda^2>1$) for $|\mathbf{k}| \ll (4\pi)^{1/2}/\lambda\approx \SI{2.4}{\per\micro\meter}$, we crop the $n_k(\mathbf{k})$ data accordingly, before Fourier transforming to $G^{(1)}(s)$. The experimentally recorded momentum distributions $n_k(\mathbf{k})$ are circularly symmetric. This agrees with the theoretical expectation, as exemplarily shown in Fig.~\ref{fig:5}(A) calculated for a normal gas with chemical potential $\mu = - \kB T$. The resolution in momentum space is $\Delta k = \SI{0.0286}{\per\micro\meter}$, and in real space we have $\Delta s = \SI{1.2}{\micro\meter}$ for $L = \{40, 50, 60\}\SI{}{\micro\meter}$ and $\Delta s = \SI{2.2}{\micro\meter}$ for $L = \SI{80}{\micro\meter}$ because the resolution in $G^{(1)}(s)$ depends on the cropping window of the momentum space images. The radially averaged data is fitted with
\begin{equation}
    g^{(1)}(s)\approx\frac{1}{\sqrt{s}} e^{-s/\xi},
    \label{eq:S6}
\end{equation}
with correlation length $\xi$ and distance $s = (s_x^2+s_y^2)^{1/2}$; eq.~\eqref{eq:S6} is valid for points spaced by $s\gg\lambda$ and describes the growth of the spatial correlations in the quantum degenerate 2D Bose gas~\cite{Hadzibabic:2011}. We therefore restrict our fit range to $s > \SI{5}{\micro\meter}$, which defines the lower fit limit. The upper fit limit is determined by the system size $L$ which limits the measurable degree of coherence at larger distances. The extent is indicated by the dashed line in Fig.~\ref{fig:2}(A) of the main text (bottom panels) and in Fig.~\ref{fig:5}(B), showing the calculated contour $(s_x,s_y)$ where the correlation function $G_0^{(1)}(s_x,s_y)=\int |\phi_{1,1}({\bf k})|^2 e^{i{\bf k}{\bf s}}d{\bf k}$ of the low-energy ground state $\phi_{1,1}({\bf k})=4\pi L [\cos(k_xL/2)\cos(k_yL/2)]^2/[(\pi^2-k_x^2L^2)(\pi^2-k_y^2L^2)]$ in the trap potential has dropped to $5\%$ of its central value $G_0^{(1)}(0,0)$. The good agreement of the calculation with the experimental data for $G^{(1)}(s_x,s_y)$ confirms that it is indeed the ground state that imposes the relevant upper limit for the range of measurable correlations. Numerically, we find that the theoretical and experimental correlation functions agree for $s\lesssim L/3$ and therefore chose it as an upper fit limit. The measurement is also partially limited by noise, and the smallest reliable measurement gives $g^{(1)}(s) = 0.007$ (see red data in Fig.~\ref{fig:2}(C) of the main text). To account for systematic effects in the fitting of $g^{(1)}(s)$, we varied the fit range of $s$ by around $20\%$ for each dataset separately. When averaging the obtained values of $\xi$ and the fit errors, the resulting relative systematic error on $\xi$ is around $20\%$.

The experimental data $G^{(1)}(s)$ [see e.g. Fig.~\ref{fig:5}(B)] show a cross-shaped feature, whose physical origin is presently not fully understood. We have excluded that the cross is an artifact of the Fourier transform data analysis. We have also verified that it originates from the microcavity geometry itself by rotating the cavity mirror about the optical axis and observing that the cross-shaped signal rotates accordingly. Therefore, the signal is most likely related to residual light scattering at the boundaries of the box potential. Importantly, we have verified that the cross does not alter the extracted critical scaling data within our experimental accuracy, by using an amplitude mask. Accordingly, we use the raw data for further analysis, especially as the underlying physical mechanism is presently not fully understood. Figure~\ref{fig:5}(B) shows raw (left) and masked data (right) and varying the masking angle yields only a $2\%$ variance in the extracted $\xi$. Figure~\ref{fig:5}(C) shows radial averages, which within the fit range show only minor deviations and we find that the extracted critical scaling and exponent is robust with respect to the masking procedure.

\subsection{4. Pöschl-Teller density of states}

To study the critical behavior, we consider the photon gas to be confined in a Pöschl-Teller potential $V(x,y) = \alpha(\alpha-1)\pi^2\hbar^2/(2mL^2)\cdot [\cos^{-2}(\pi x/L) + \cos^{-2}(\pi y/L)]$~\cite{Poeschl:1933,Nieto:1978}. This potential allows one to take into account rounded edges at the bottom of the box trap. At low energies, which is the relevant regime for the existence of a BEC in the thermodynamic limit, our experimentally realized trap potentials are more closely described by a Pöschl-Teller potential than by a rigid box potential. In general, the expression for the Pöschl-Teller potential interpolates the Hamiltonian between that one of a photon gas inside a rigid box potential ($\alpha=1$) and inside a harmonic trap ($\alpha,L\rightarrow\infty$), for which the trap frequency $\Omega$ is determined by the ratio $\alpha/L^2 = m \Omega /(\pi^2\hbar)$. In this section, we derive the density of states in this potential.

\begin{figure}[t]
  \centering
  \includegraphics[width=1\columnwidth]{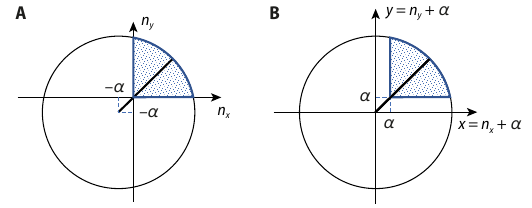}
 \caption{$\vert$ \textbf{State space in the Pöschl-Teller potential. (A)} Graphical representation of the number of states $\Gamma(\epsilon)$ given in eq.~\eqref{eq:S9}. The number of states corresponds to the blue shaded section of a displaced circle, as obtained from the energy spectrum in the Pöschl-Teller potential. \textbf{(B)} $\Gamma(\epsilon)$ is obtained by integrating the displaced circle.}
\label{fig:6}
\end{figure}

\begin{figure*}
  \centering
  \includegraphics[width=0.7\textwidth]{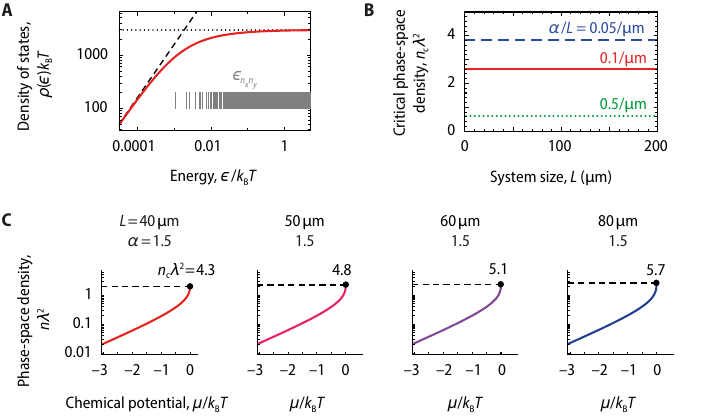}
 \caption{$\vert$ \textbf{Density of states and equation of state in the Pöschl-Teller potential. (A)} Density of states from eq.~\eqref{eq:S10} as a function of energy for $\alpha=1.5$ and $L = \SI{80}{\micro\meter}$. The corresponding eigenstates from eq.~\eqref{eq:S7} are shown as gray solid lines. \textbf{(B)} Critical phase-space density as a function of box size $L$, for three values of $\alpha/L = \{0.05,0.1,0.5\}$. The critical value is constant for $\alpha/L = \mathrm{const.}$ \textbf{(C)} Equation of state in the Pöschl-Teller potential for the experimental $\alpha = 1.5$ and $L = \{40, 50, 60, 80\}\SI{}{\micro\meter}$. At $\mu=0$, the critical phase-space density is reached, and a phase transition is expected. As in the experiments, the ratio $\alpha/L$ varies between different panels. Correspondingly, the critical phase-space density becomes larger.}
\label{fig:7}
\end{figure*}

The density of states can be obtained from the energy spectrum of the Pöschl-Teller potential in two dimensions~\cite{Nieto:1978},
\begin{equation}
    \epsilon_{n_x,n_y} = \frac{\hbar^2}{2m}\left(\frac{\pi}{L}\right)^2 \left[ (n_x+\alpha)^2 +(n_y+\alpha)^2 \right]
    \label{eq:S7}
\end{equation}
with the quantum numbers $n_x, n_y = 0,1,2,...$ and so on. For convenience, the excitation energy can be measured from the ground state energy as $\epsilon\equiv \epsilon_{n_x,n_y}- \epsilon_{0,0}$ where $\epsilon_{0,0} = 2V_0$ and $V_0 = \hbar^2/(2m) (\pi\alpha/L)^2$. This way, the chemical potential takes negative values or zero only. The density of states is given by
\begin{equation}
    \rho(\epsilon) = \frac{d\Gamma(\epsilon)}{d\epsilon}
    \label{eq:S8}
\end{equation}
where $\Gamma(\epsilon)$ is the number of states with energy less or equal than $\epsilon$. In the thermodynamic limit, $\Gamma(\epsilon)$ corresponds to the shaded area indicated in Fig.~\ref{fig:6}(A). This is because the energy spectrum can be accordingly written as $\tilde \epsilon + 2\alpha^2 = (n_x + \alpha)^2 + (n_y + \alpha)^2$ with $\tilde \epsilon = 2m/\hbar^2 (L/\pi)^2 \epsilon$, which is the equation of a circle centered at $(-\alpha,-\alpha)$. For easiness of calculation, we make a change of variables $x = n_x + \alpha, y = n_y + \alpha$, and $\Gamma(\epsilon)$ can be obtained by integrating the blue shaded area in Fig.~\ref{fig:6}(B). This is, $\Gamma(\epsilon) = \int_\alpha^{\sqrt{\tilde \epsilon+\alpha^2}} {dx} \int_\alpha^{\sqrt{\tilde \epsilon+2\alpha^2-x^2}} {dy}$, which explicitly yields,
\begin{widetext}
\begin{equation}
    \Gamma(\epsilon) = \frac12 (\tilde \epsilon + 2\alpha^2) \left( \arctan\sqrt{\frac{\tilde\epsilon + \alpha^2}{\alpha^2}} - \arctan \sqrt{\frac{\alpha^2}{\tilde \epsilon + \alpha^2}} \right) - \alpha \sqrt{\tilde\epsilon + \alpha^2} +\alpha^2
    \label{eq:S9}
\end{equation}
By differentiation with respect to $\epsilon$, see eq.~\eqref{eq:S8}, the density of states is obtained:
\begin{equation}
    \rho(\epsilon) = \frac{m}{\hbar^2} \left(\frac{L}{\pi}\right)^2 \left(\arctan\sqrt{\frac{\epsilon + V_0}{V_0}} - \arctan\sqrt{\frac{V_0}{\epsilon + V_0}}\right).
    \label{eq:S10}
\end{equation}
\end{widetext}
Figure~\ref{fig:7}(A) shows a plot of the density of states in log-log representation, indicating that the expression in eq.~\eqref{eq:S10} is governed by two regimes: At small energies, $\rho(\epsilon)\sim \epsilon$ scales linearly, and therefore resembles the density of states in a harmonic oscillator potential. In the intermediate regime, as is relevant for the low-energy states of the here used box potentials [see Fig.~\ref{fig:7}(A)], the density of states increases with energy with a positive but gradually decreasing slope, while for large energies $\rho(\epsilon)\sim m L^2/(\hbar^2\pi^2)$ becomes constant and approaches the expectations of a homogeneous system.

\subsection{5. Divergence of the correlation length}

In this section we analytically derive an expression of the critical scaling behavior of the correlation length of the form $\xi(t)\sim|t|^{-\nu}$ for the photon gas near the phase transition at $t = 0$. For this, we again analyze the Pöschl-Teller potential~\cite{Poeschl:1933,Nieto:1978}. Using the expressions from Section 4, the equation of state for the phase-space density
\begin{equation}
    \mathcal{D}(\mu)=n(\mu)\lambda^{2} = \bar g (\mu/\kB T, V_0/\kB T),
    \label{eq:S11}
\end{equation}
follows from the generalized Bose function $\bar g (\mu/\kB T, V_0/\kB T)= \int_0^\infty\, dz\, \bar\rho(z,V_0/\kB T)/[\exp(z-\mu/\kB T)-1]$. The expression $\bar\rho(z,V_0/\kB T)=2\pi\hbar^2/(m L^2)\cdot\rho(\epsilon)|_{\epsilon=z\kB T}$ results from the density of states in the Pöschl-Teller potential in eq.~\eqref{eq:S10}, such that Bose-Einstein condensation in 2D at $\mu=0$ is formally possible at a finite critical particle number
\begin{equation}
    \Nc=\left(\frac{L}{\lambda}\right)^{2} \bar g (0, V_0/\kB T).
    \label{eq:S12}
\end{equation}

The existence of Bose-Einstein condensation at finite temperature can be understood from the, as compared to the rigid box case, reduced density of states at low energies, see also the above discussion. When approximating eq.~\eqref{eq:S11} for small $\mu\to0^-$, one has $\mathcal{D}(\mu)\approx \Dc-\mu/\pi V_0\ln(-\mu/\kB T)$. Note that the value of the critical phase-space density $\Dc$ is well defined and yields a finite critical temperature $T_c=\textrm{const.}$ when taking the appropriate thermodynamic limit ($N$, $L\to\infty$ with $\alpha/L=\textrm{const.}$) which conserves the nearly uniform trapping also for $L\to\infty$. In contrast, in the limit $L\to\infty$ with $\alpha=\textrm{const.}$, also $\Dc$ goes to infinity indicating that Bose-Einstein condensation is suppressed, as expected in the infinite 2D homogeneous case. Figure~\ref{fig:7}(B) shows the critical phase-space density obtained by numerical integration as a function of the system size $L$ for three different ratios of $\alpha/L$. The plot demonstrates that the critical value remains conserved even when increasing $L$ (and adjusting $\alpha$ accordingly). The equation of state $\mathcal{D}(\mu)$ for the four sets of experimental parameters with $\alpha=1.5=\textrm{const.}$ and increasing $L$ are shown in Fig.~\ref{fig:7}(C). Here the ratio $\alpha/L$ is not maintained, and consequently the critical phase space density grows with $L$, in agreement with the experimental results.

Using the relation $t=(T-T_c)/T_c=(N_c-N)/N=(\Dc-\mathcal{D})/\mathcal{D}\simeq(\Dc-\mathcal{D})/\Dc$ for small $t$ and inserting $\mu=-\hbar^2/(2m\xi^2)$, which follows from the definition of the correlation length $\xi=\hbar/(2m|\mu|)^{1/2}$. The implicit exact expression of the correlation length is derived from the Bose-Einstein distribution in the quantum degenerate limit $n_k\approx1/[k^2+(1/\xi)^2]$, see e.g. Ref.~\cite{Hadzibabic:2011}
\begin{equation}
    \xi^{-2}\ln\left(\sqrt{4\pi} \frac{\xi}{\lambda}\right)\approx C_0 t
    \label{eq:S13}
\end{equation}
with $C_0=2\pi^2V_0n_c/\kB T$ a constant. Equation~\eqref{eq:S13} can be iteratively solved for $t$ in the asymptotic limit $t\to 0^+$, yielding an explicit expression for the divergence of the correlation length at the critical point,
\begin{equation}
    \xi\approx \left[\frac{1}{2C_0 } \frac{\ln (1/t)}{t}\right]^{1/2}
    \label{eq:S14}
\end{equation}
hinting at the algebraic scaling with a critical exponent $\nu \approx 1/2$ and logarithmic corrections. A similar algebraic scaling form can also be derived for the harmonically trapped 2D Bose gas~\cite{Morales-Amador:2024}. 

\begin{figure}[t]
  \centering
  \includegraphics[width=1\columnwidth]{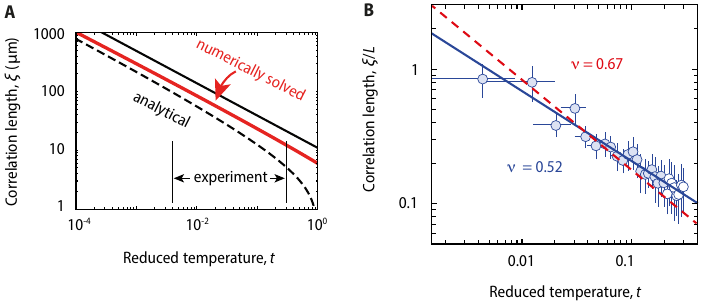}
 \caption{$\vert$ \textbf{Scaling behavior. (A)} Numerical solution of the exact expression eq.~\eqref{eq:S13} (red line), which is well described by power law (black line) $\sim t^{-0.55}$ (vertically shifted for comparison), along with approximate analytical expression from eq.~\eqref{eq:S14} (dashed line). The latter approaches the exact behavior only for small $t$, smaller than in our experiments, as indicated by vertical lines. \textbf{(B)} Experimental data for $L = \SI{50}{\micro\meter}$ (circles), see Fig.~\ref{fig:4}(A) of the main text, along with fitted power law for the ideal Bose gas (solid blue line); for comparison, the power law with $\nu = 0.67$ expected for an interacting Bose gas in the 3D XY model class is shown (dashed red line). The error bars show combined systematic and statistical uncertainties.}
\label{fig:8}
\end{figure}

Directly fitting the experimental data with the approximate eq.~\eqref{eq:S14}, however, is not valid in the studied range of reduced temperatures $0.004 < t < 0.3$. In this window, the $\ln(1/t)$ term causes artificial distortions of $\nu = -\text{d}(\ln \xi)/\text{d}(\ln t)$, because the asymptotic limit is not met. This is seen by numerically evaluating eq.~\eqref{eq:S13}, see the red line in Fig.~\ref{fig:8}(A), which to good approximation shows a power law with $\nu = 0.55(1)$ (black solid line), while eq.~\eqref{eq:S14} (dashed line) fails to adequately describe the scaling behavior. The results shown in Fig.~\ref{fig:8}(A) provide an important justification why fitting a power law function to the data is appropriate for our study and that logarithmic effects indeed are weak.

\subsection{6. Error estimation for measured critical exponent}

The present work reports the critical exponent $\nu = 0.52(4)$ for a trapped ideal Bose gas near the BEC phase transition, which within experimental uncertainties differs from the exponent for the interacting Bose gas $\nu = 0.67$, indicating their distinct universality classes, see Fig.~\ref{fig:8}(B). This section discusses our analysis of systematic and statistical uncertainties for the critical exponent.

\vspace{-0.4cm}
\subsubsection{Systematic uncertainties} 
\vspace{-0.2cm}

We first demonstrate the robustness of $\nu$ against systematic effects. For this, we have varied the fit range of reduced temperatures $t$ and taken into account the uncertainty in the critical particle number $\Nc$. We repeat the fit for all sub-ranges that span at least one decade in the reduced temperature between $0 < t < 0.25$. The systematic uncertainty (for each system size) is then determined as the standard deviation $\sigma_{\text{sys,range}}$ over the resulting exponents, see Table~\ref{tab:final_results}. To assess the influence of the experimental uncertainty of $\Nc$ on the critical exponent $\nu$, we re-evaluate the reduced temperatures $t$ and repeat the fit. Because $\Delta\Nc$ propagates into $t = (\Nc–N)/\Nc$, orthogonal distance regression is employed for all critical scaling fits to account for uncertainties in $\ln(\xi)$ and $\ln(t)$. The systematic uncertainty related to $\Nc$ follows from the mean absolute deviation $\sigma_{\text{sys}, \Nc} = ( |\nu(\Nc+\Delta \Nc)-\nu(\Nc)| + |\nu(\Nc-\Delta \Nc)-\nu(\Nc)| )/2$. The total systematic errors $\sigma_{\text{sys,tot}} = (\sigma_{\text{sys,range}}^2 + \sigma_{\text{sys}, \Nc}^2)^{1/2}$ are summarized in Table~\ref{tab:final_results}.

\begin{table*}
\centering
\begin{tabular}{|c||c||c|c|c||c|c||c|c|}
    \hline
    \js{Dataset} & \js{$\nu$} & \js{$\sigma_{\text{sys}, \text{range}}$} & \js{$\sigma_{\text{sys}, N_c}$} & \js{$\sigma_{\text{sys}, \text{tot}}$} & \js{$\sigma_{\text{stat, low}}$} & \js{$\sigma_{\text{stat, high}}$} & \js{$\sigma_{\text{tot, low}}$} & \js{$\sigma_{\text{tot, high}}$} \\ \hline
    \js{40\,µm} & \js{0.406} & \js{0.102} & \js{0.082} & \js{0.1309} & \js{0.077} & \js{0.070} & \js{0.1518} & \js{0.1484} \\ \hline
    \js{50\,µm} & \js{0.521} & \js{0.012} & \js{0.029} & \js{0.0314} & \js{0.054} & \js{0.049} & \js{0.0625} & \js{0.0582} \\ \hline
    \js{60\,µm} & \js{0.558} & \js{0.081} & \js{0.034} & \js{0.0878} & \js{0.057} & \js{0.076} & \js{0.1047} & \js{0.1162} \\ \hline
    \js{80\,µm} & \js{0.514} & \js{0.029} &\js{0.056} & \js{0.0631} & \js{0.055} & \js{0.043} & \js{0.0837} & \js{0.0763} \\ \hline
\hline
\multicolumn{2}{|c|}{\textbf{\js{Weighted Mean}}} & \multicolumn{7}{c|}{\js{$\mathbf{\bar{\nu} = 0.52 \pm 0.04}$}} \\ \hline
\end{tabular}
\caption{$\vert$ \textbf{Experimental results and error budget for the critical exponent $\nu$.} Summarized results for the individual data sets with different box sizes and the corresponding systematic, statistical, and total uncertainties. Note that $\sigma_{\text{tot, low}} = (\sigma_{\text{sys, tot}}^2 + \sigma_{\text{stat, low}}^2)^{1/2}$ and correspondingly for $\sigma_{\text{tot, high}}$.}
\label{tab:final_results}
\end{table*}

\vspace{-0.4cm}
\subsubsection{Statistical uncertainties} 
\vspace{-0.2cm}

The statistical uncertainty is extracted from a bootstrapping method with $5000$ iterations per dataset. For this, the raw data ($\ln t$, $\ln \xi$) are randomly resampled and fitted with $f(t) = a_0 t^{–\nu}$, yielding a histogram with statistical uncertainties extracted from the $16$th and $84$th percentiles, corresponding to a $1\sigma$-confidence interval. This approach also captures asymmetries in the distribution. The corresponding values for $\sigma_{\text{stat,low/high}}$ are given in Table~\ref{tab:final_results}.

\vspace{-0.4cm}
\subsubsection{Total uncertainties} 
\vspace{-0.2cm}
Combining the systematic and statistical uncertainties per data set provides the overall error estimate for the critical exponent. We obtain the weighted mean $\bar{\nu}=(\sum_{i=1}^4 {w_i\nu_i})/\sum_{i=1}^4 w_i = 0.52$, with weights $w_i = 4/(\sigma_{\text{stat,low}}+\sigma_{\text{stat,high}})^2$. The error is calculated as $\sigma_\nu = 1/(\sum_{i=1}^{4}w_i)^{1/2}\ = 0.04$. Within error bars $\bar{\nu}$ is consistent with the individual $\nu_i$ for the different system sizes, and the results confirm that that the obtained critical exponents for all data sets differ from $\nu = 0.67$ within uncertainties. Figure~\ref{fig:8}(B) shows the experimental data for $L = \SI{50}{\micro\meter}$ which is well described by a power law with the fitted exponent $\nu=0.52$ (solid blue line). In contrast, a power law with $\nu = 0.67$ (dashed red line), as for an interacting Bose gas in the XY model universality class visibly deviates from the experimental data. The experimental data for the other box sizes shows a similar degree of disagreement with the $\nu = 0.67$ XY model universality class prediction. For a summary of the results of the critical exponent in the data sets taken for four different box sizes, as well as our averaged final result, see also Table~\ref{tab:final_results}.

\vspace{-0.4cm}
\subsubsection{Data point sampling} 
\vspace{-0.2cm}

The inset of Fig.~\ref{fig:4}(A) in the main text shows that the data point density between reduced temperatures $t = 0.05$ and $0.2$ is higher than at smaller $t$. To exclude that the fit is dominated by the data at large $t$, we have performed a consistency check using logarithmic binning. For each bin we calculate the mean values of $t$ and $\xi$, and the mean statistical error. This procedure leads to a uniform data point density across the scaling range. A weighted least-squares regression of the binned mean points gives $\nu = 0.48(6)$, which within uncertainties agrees with the unbinned results and confirms that the scaling behavior is not an artifact of the point density.

\end{document}